\documentclass[sigconf,10pt]{acmart}

\AtBeginDocument{%
  \providecommand\BibTeX{{%
    \normalfont B\kern-0.5em{\scshape i\kern-0.25em b}\kern-0.8em\TeX}}}

\setcopyright{acmlicensed}
\copyrightyear{2024}
\acmYear{2024}
\acmDOI{XXXXXXX.XXXXXXX}

\acmConference[Conference acronym 'XX]{Make sure to enter the correct
  conference title from your rights confirmation email}{June 03--05,
  2018}{Woodstock, NY}
\acmISBN{978-1-4503-XXXX-X/18/06}

\usepackage[linesnumbered,ruled,algoruled,boxed,lined]{algorithm2e}
\usepackage{algorithmic}
\usepackage{tikz}
\usetikzlibrary{positioning,shapes,arrows,arrows.meta}
\usepackage{amsmath}
\usepackage{listings}
\usepackage{tabularx}
\usepackage{longtable}
\usepackage{paralist}
\usepackage{xspace}
\usepackage{adjustbox}
\usepackage{multirow}
\usepackage{footnote}
\usepackage{xurl}
\usepackage{booktabs}
\usepackage{float}
\usepackage{subcaption}
\usepackage{listings}
\usepackage{xcolor,colortbl}
\usepackage{hyperref}
\usepackage{color}
\usepackage{colortbl}
\usepackage{multirow}
\usepackage{makecell}
\usepackage{enumitem}
\hypersetup{colorlinks=false, pdfborder={0 0 0}}
\usepackage{array}
\usepackage{seqsplit}

\newcommand{\ie}{i.\@\,e.\@\xspace}
\newcommand{\eg}{e.\@\,g.\@\xspace}

\newcommand{\ourname}{Dyna-5G\xspace}

\lstdefinestyle{customstyle}{
    basicstyle=\ttfamily\small,
    breaklines=true,
    frame=tb,
    xleftmargin=.05\textwidth, 
    xrightmargin=.05\textwidth,
    keywordstyle=\color{blue},
    commentstyle=\color{red},
    morekeywords={ENTRY_NOTIFICATION, Fields, PERFORMANCE_REPORT, Description, field, message, messages, mavlink, EXIT_NOTIFICATION, LOCATION_UPDATE_NOTIFICATION, HEARTBEAT_NOTIFICATION, MISSION_PLAN_NOTIFICATION, ENTRY_NOTIFICATION_REPLY},
}

\definecolor{headercolor}{rgb}{0.92,0.92,0.99} 
\definecolor{rowcolorA}{rgb}{0.98,0.98,0.98} 
\definecolor{rowcolorB}{rgb}{1,1,1} 

\begin{document}

\title{Dyna-5G: Dynamic Role Switching for Self-Organizing 5G M2M Networks}

\author{Evangelos Bitsikas}
\email{bitsikas.e@northeastern.edu}
\affiliation{%
  \institution{Northeastern University}
  \city{Boston}
  \state{MA}
  \country{USA}
}

\author{Adam Belfki}
\email{belfki.a@northeastern.edu}
\affiliation{%
  \institution{Northeastern University}
  \city{Boston}
  \state{MA}
  \country{USA}
}

\author{Aanjhan Ranganathan}
\email{aanjhan@northeastern.edu}
\affiliation{%
  \institution{Northeastern University}
  \city{Boston}
  \state{MA}
  \country{USA}
}

\begin{abstract}

M2M deployments such as drone swarms demand mission-critical communication: km-scale range, strong per-device identity and mutual authentication, and deterministic QoS for bandwidth-intensive payloads. Cellular 5G uniquely satisfies all of these — yet it has seen limited adoption in autonomous fleets. The barrier is not capability but resilience: today's 5G networks assume fixed infrastructure, and when the base station fails, recovery is uniquely complex. Unlike simpler wireless protocols where devices can transparently switch nodes, 5G failure requires reconstructing distributed state such as authentication contexts, QoS bindings, tunnels, and RRC state machines across the fleet, a process that no existing system automates.
We present \ourname, which makes this happen. \ourname is the first 5G Standalone-compliant framework for dynamic role switching in M2M fleets, where any device can assume the role of 5G Core, RAN, or UE at runtime. It orchestrates failure detection, leader selection, and coordinated state teardown and re-establishment, all without modifying 3GPP protocols. 
We evaluate \ourname on a high-fidelity software emulation testbed, with Open5GS and srsRAN, across hundreds of trials with up to 10 drones. Control-plane overhead averages 0.47~Mb/s (${\approx}$0.47\% of a 100~Mb/s bearer), while failure recovery completes in ${\approx}$2.5~s, of which ${\approx}$86\% is due to stack-dependent cellular procedures. 
\ourname's orchestration logic itself adds only ${\approx}$175~ms per reconfiguring role. All tested missions complete successfully, even under injected leader crashes.

\end{abstract}



\maketitle

\section{Introduction}


Large machine-to-machine (M2M) applications, including drone swarms for emergency response, autonomous vehicle fleets, and smart factory coordination, demand mission-critical wireless connectivity: kilometer-scale range, strong per-device identity and mutual authentication, deterministic quality-of-service for bandwidth-intensive payloads, and the ability to connect thousands of devices simultaneously. Cellular 5G NR uniquely satisfies all of these through licensed spectrum, SIM-based authentication, and granular QoS policies~\cite{5gamericas1, 3gpp.38.913, ericsson22:5g, qualcomm22:5g-scaling}. Yet in practice, many M2M deployments still rely on proprietary solutions or ad-hoc mesh networks, often favoring deployment simplicity over tighter integration with cellular infrastructure. This raises a natural question: can 5G NR be deployed directly within an M2M fleet, without dependence on fixed cellular infrastructure?

The foremost challenge is resilience. Autonomous fleets rely on continuous, distributed coordination algorithms~\cite{Wang22:consensus-multi-graph, Chen20:drone-consensus, Chen24:drone-coordination, Ozmen23:ShortRS, Alsamhi23:blockchain-drone, Campion18:drone-communication, Schwager08:ConsensusLF, Schwager:11UnifyingGP} (le\-ader-follower, consensus, swarm coordination) that cannot tolerate network partitions; pausing to wait for human intervention is not an option mid-mission. Today's 5G networks assume fixed infrastructure, and when the base station fails, the entire fleet loses connectivity, a classic single point of failure. But recovering from this is far more complex than simply electing a new leader: when the drone hosting the 5G Core and RAN crashes mid-mission, every node in the fleet must tear down and rebuild its entire connection state with the newly elected leader, in a coordinated sequence, before the network can resume. Unlike simpler wireless protocols where devices can transparently switch access points, 5G failure requires reconstructing tightly coupled distributed state that spans the Core, RAN, and each UE simultaneously. No existing system automates this. 
Beyond failure recovery, static 5G is architecturally misaligned with how M2M fleets already operate: the distributed coordination algorithms they run assume nodes can exchange roles dynamically, a capability that fixed infrastructure cannot provide.
In this paper, we explore how to make it happen.

In this work, we present \textbf{\ourname}, the first 5G Standalone (SA)-compliant framework for dynamic role switching in M2M fleets. In \ourname, any machine can assume the role of 5G Core, RAN, or User Equipment (UE) at runtime. A state-machine controller on each device drives initialization, leader selection, network entry and exit, and failure recovery, orchestrating coordinated state teardown and re-establishment across the fleet without modifying any 3GPP protocol. Role assignment is algorithm-agnostic: a scoring module can be replaced to customize behavior for different missions and environments without altering the rest of the system.


Specifically, we make the following contributions:

\begin{enumerate}
       \item We design and implement \textbf{\ourname}, the first 5G SA-compliant framework for dynamic role switching in M2M fleets. Its core is a distributed state-machine controller that coordinates role transitions across all nodes, covering initialization, leader selection, network entry, exit, and failure recovery, using standard 3GPP procedures without any protocol modification.

    \item We design and implement failure recovery mechanisms that automate coordinated state reconstruction across the fleet, detecting failures via missed heartbeats and cellular state inspection, and autonomously triggering leader election and node removal to preserve mission continuity.

    \item We validate system correctness and scalability using a high-fidelity software emulation pipeline with Open5GS and srsRAN, replaying simulated missions with up to 10 drones across hundreds of trials. Control-plane overhead averages $\approx$0.47~Mb/s at 10 drones (98-byte packets), application capacity reaches tens of Mb/s TCP goodput with zero retransmissions and lossless UDP at 16~Mb/s with sub-ms jitter, and end-to-end latencies remain in the tens of milliseconds.

    \item We demonstrate robust adaptation under failures by injecting leader and follower crashes mid-mission. \ourname detects loss, removes the failed node, and completes emergency leader selection autonomously. The orchestration logic itself adds only $\approx$175~ms per role reconfiguration, with total recovery completing in $\approx$2.5~s, of which $\approx$86\% is attributable to standard cellular stack procedures. All tested missions complete successfully.
\end{enumerate}

\section{Motivating Scenario and Related Work}



\begin{figure}[!t]
     \centering
     \includegraphics[width=0.9\columnwidth]{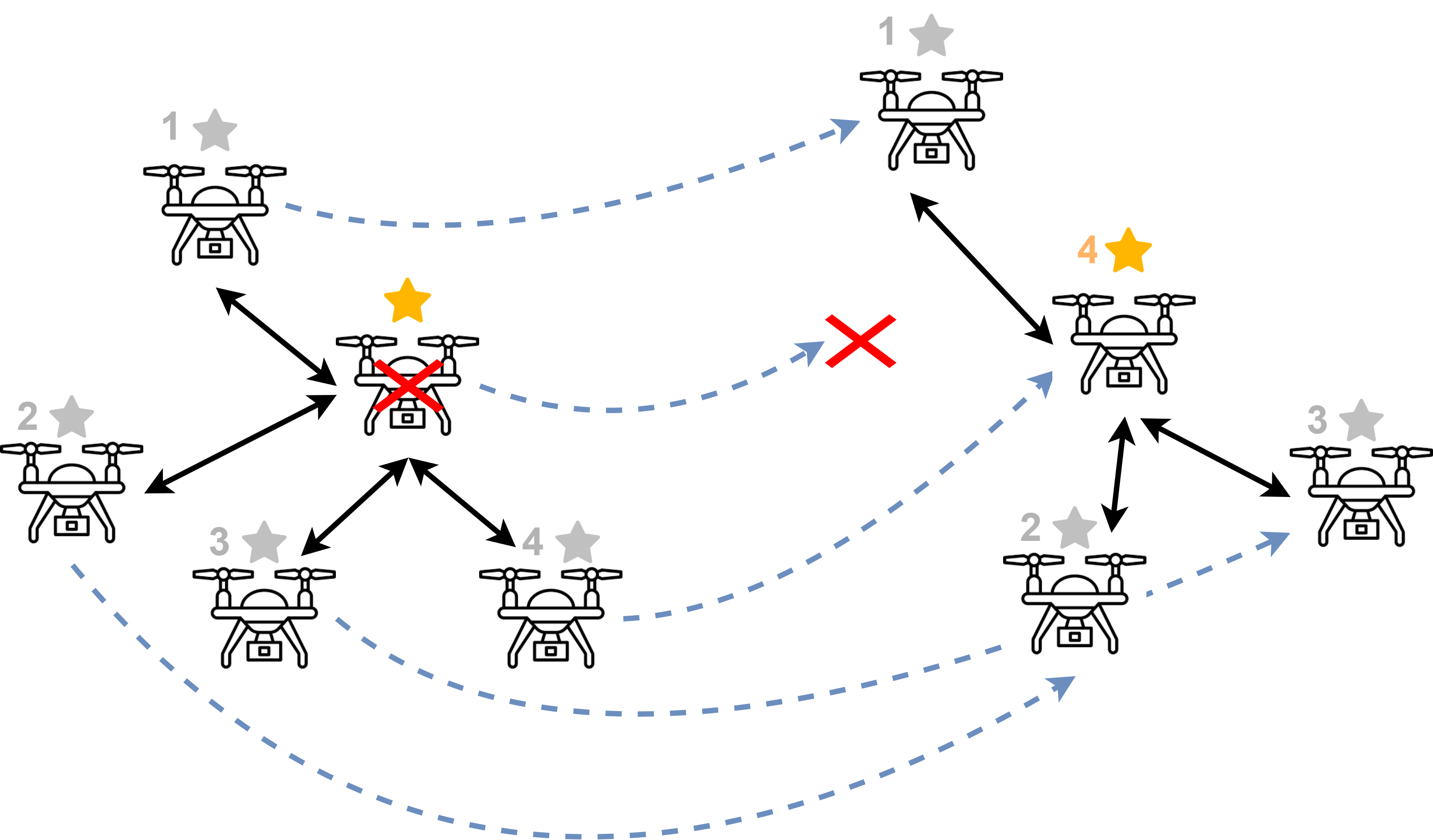}
     \caption{The drones execute a mission (blue shows the trajectory), when the leader fails unexpectedly. The network reorganizes with a new leader (\#4).
     }
     \Description{}
     \label{fig:drone-networks}
\end{figure}

We use a drone swarm as a running example throughout this paper. Figure~\ref{fig:drone-networks} illustrates the core scenario: drones execute a mission until the leader (the drone hosting the 5G Core and RAN) fails unexpectedly, triggering a network reorganization. We next explain why existing wireless approaches cannot adequately address this scenario.

\subsection{Why Not 5G NR Sidelink or Ad-Hoc?}

NR sidelink~\cite{3gpp.38.321, 3gpp.38.331} is an important alternative when the goal is direct UE-to-UE communication with reduced dependence on the base station for the data path. However, it addresses a different problem from Dyna5G. Our goal is not simply to bypass the gNB, but to explore whether a fleet can preserve cellular-style operation by dynamically re-instantiating infrastructure roles within the fleet itself. This matters when the mission requires capabilities associated with infrastructure-managed service, such as session anchoring, centralized policy and QoS control, SIM-based security contexts, and compatibility with applications that assume gNB/5GC-provided connectivity. We therefore view sidelink/ad-hoc approaches and Dyna5G as addressing different system objectives: the former focuses on infrastructure-light direct communication, whereas the latter focuses on dynamic reorganization of the cellular infrastructure itself.

\subsection{Related Work}


There is extensive work on enhancing LTE/5G legacy infrastructure~\cite{Cho18:mobilestream, ChenK18:d2d, popescu21:connecting, Anand22:decentralized, Ahmad23:Ahmad, Moradi18:SoftBox, Mozaffari16:Unmanned-d2d, Jin13:softcell}, including core and RAN decentralization and device-to-device communication. Drone-assisted cellular networks~\cite{Sobouti20:smallcell, Singhal21:aerialson, Bekkouche20:service-based, Abdalla21:3gpp, Sekander18:drone-5g, Koumaras21:5g-uav} have been explored to extend coverage or capacity, along with specialized testbeds~\cite{aerpaw23:Wireless-testbed, bonati21:colosseum} for experimentation. Surveys~\cite{Wu21:Overview-5G, hakak2022autonomous} assess the current state of vehicular and aerial 5G deployments. These efforts improve existing infrastructure but treat the cellular architecture as fixed: core, RAN, and UE roles are statically assigned and not subject to runtime switching.


Closest to our goals are systems that place LTE network functions on airborne nodes~\cite{sheshadri2020skyhaul, Chakraborty2018SkyRANAS, Moradi2018SkyCoreMC, Sundaresan2018SkyLiTEED}. SkyCore~\cite{Moradi2018SkyCoreMC} deploys an edge EPC on UAVs to reduce dependence on terrestrial core infrastructure. SkyRAN~\cite{Chakraborty2018SkyRANAS} mounts an LTE RAN on a drone for dynamic coverage. SkyHaul~\cite{sheshadri2020skyhaul} builds an autonomous UAV backhaul using mmWave mesh links. SkyLiTE~\cite{Sundaresan2018SkyLiTEED} augments LTE connectivity in sparse-infrastructure areas. While these systems show that cellular functions can be hosted on mobile nodes, they do not support role switching: a drone acting as base station remains the base station, with no mechanism to elect a replacement or reconstruct cellular state when it fails. Furthermore, these systems operate on monolithic LTE architectures for coverage area scenarios only; moving to 5G Standalone introduces complex, distributed state management across its Service-Based Architecture. None of the existing work automates the coordinated state teardown based on thorough failure recovery and re-establishment required to transition roles at runtime. \ourname is the first to automate dynamic role transitions and state reconstruction for 5G Standalone networks.

\section{Design of \ourname}


\begin{table*}[!t]
    \centering
    \caption{FSM Actions and Transitions}
    \label{tab:fsm_transitions}
    \begin{tabular}{lcll}
        \arrayrulecolor{black}
        \toprule
        \rowcolor{headercolor}
        \textbf{Action} & \textbf{Event Symbol} & \textbf{Fulfilled Condition} & \textbf{State Transition} \\
        \midrule
        \rowcolor{rowcolorA}
        Initialization & \( E_{I} \) & \( P_L(m_i) = \frac{1}{N}, P_F(m_i) = \frac{N-1}{N}\) & \( S(m_i) \xrightarrow[E_{I}]{P(m_i)} \text{Leader/Follower} \) \\
        \rowcolor{rowcolorB}
        Entering & \( E_{E} \) & $m_i$ sends \texttt{Entry Notification} & \( S(m_i) \xrightarrow[E_{E}]{} \text{Follower} \) \\
        \rowcolor{rowcolorA}
        Leader Selection & \( E_{LS} \) & \( m_{L} = \text{argmax}_{m_i \in M} S^{*}(m_i) \) & 
        \( S(m_i) \xrightarrow[E_{LS}]{m_i = m_{L}} \text{Leader} \), \( S(m_i) \xrightarrow[E_{LS}]{m_i \neq m_{L}} \text{Follower} \) \\
        \rowcolor{rowcolorB}
        Exiting & \( E_{X} \) & $m_i$ sends \texttt{Exit Notification} & \( S(m_i) \xrightarrow[E_{X}]{} \text{Null} \) \\
        \rowcolor{rowcolorA}
        Failure & \( E_{F} \) & $m_i$ exits without \texttt{Exit Notification} & \( S(m_i) \xrightarrow[E_{F}]{} \text{Null} \) \\
        \rowcolor{rowcolorB}
        Recovery & \( E_{R} \) & $m_i$ sends \texttt{Entry Notification}~\footnotemark & \( S(m_i) \xrightarrow[E_{R}]{} \text{Follower} \) \\
        \bottomrule
    \end{tabular}
\end{table*}

\subsection{Design Requirements}

Based on Figure~\ref{fig:drone-networks}, we identify four primary design requirements essential for crafting a dynamic, self-organizing 5G network that is suitable for M2M networks. These requirements are not met by conventional deployments due to the static nature of 5G's Service-Based Architecture and the tight coupling of distributed state across the Core, RAN, and UE.

\noindent \textbf{$\bullet$ Autonomous Network Organization:} Devices must be able to autonomously establish a private 5G network that does not rely on external operators or the Internet. This is essential in remote or disrupted areas, where traditional infrastructure is unavailable or unreliable. For example, in a natural disaster, rescue drones and vehicles must form a working network immediately, without waiting for public 5G coverage. Achieving this autonomy requires local control over both network decisions and cellular operations, which conventional architectures do not typically expose.

\noindent \textbf{$\bullet$ UE-UE Communication:} The network must allow UEs to communicate directly to exchange data. In a private, isolated 5G SA deployment without external infrastructure, UE-to-UE traffic must be routed locally within the fleet rather than forwarded to an unreachable upstream (as also seen in Sections~\ref{conventional-net} of Appendix), a routing configuration that standard 5G deployments do not expose by default.
Our design therefore requires careful configuration of routing and tunneling through the Core/RAN, while remaining agnostic to the machine type and avoiding extra overlay protocols.

\noindent \textbf{$\bullet$ Dynamic Core \& RAN Selection:} Machines must be free to assume UE, Core, or RAN roles dynamically, so the network can reconfigure under changing demands and conditions. This breaks with traditional 5G deployments, where Core and RAN roles are fixed and UEs remain passive. In scenarios like disaster recovery, allowing drones or emergency vehicles to temporarily act as Core or RAN nodes keeps communication running even when pre-existing infrastructure fails. Switching roles, however, requires automating the coordinated teardown and re-establishment of UE contexts, QoS bindings, and radio links across the entire fleet.

\noindent \textbf{$\bullet$ Enhanced Network Resilience:} A dynamic network must remain operational under failures, especially in mission-critical settings. Conventional networks often depend on single core components, centralized databases and rigid QoS policies, creating clear single points of failure. Instead, we require a distributed FSM that rapidly detects failures, triggers leader election, and coordinates state reconstruction across the fleet so the network continues to function when individual nodes fail. Distributing credentials and states further reduces the impact of any single compromised or unreachable component, improving overall integrity and availability.

\begin{figure}[!t]
    \centering
    \includegraphics[width=\columnwidth]{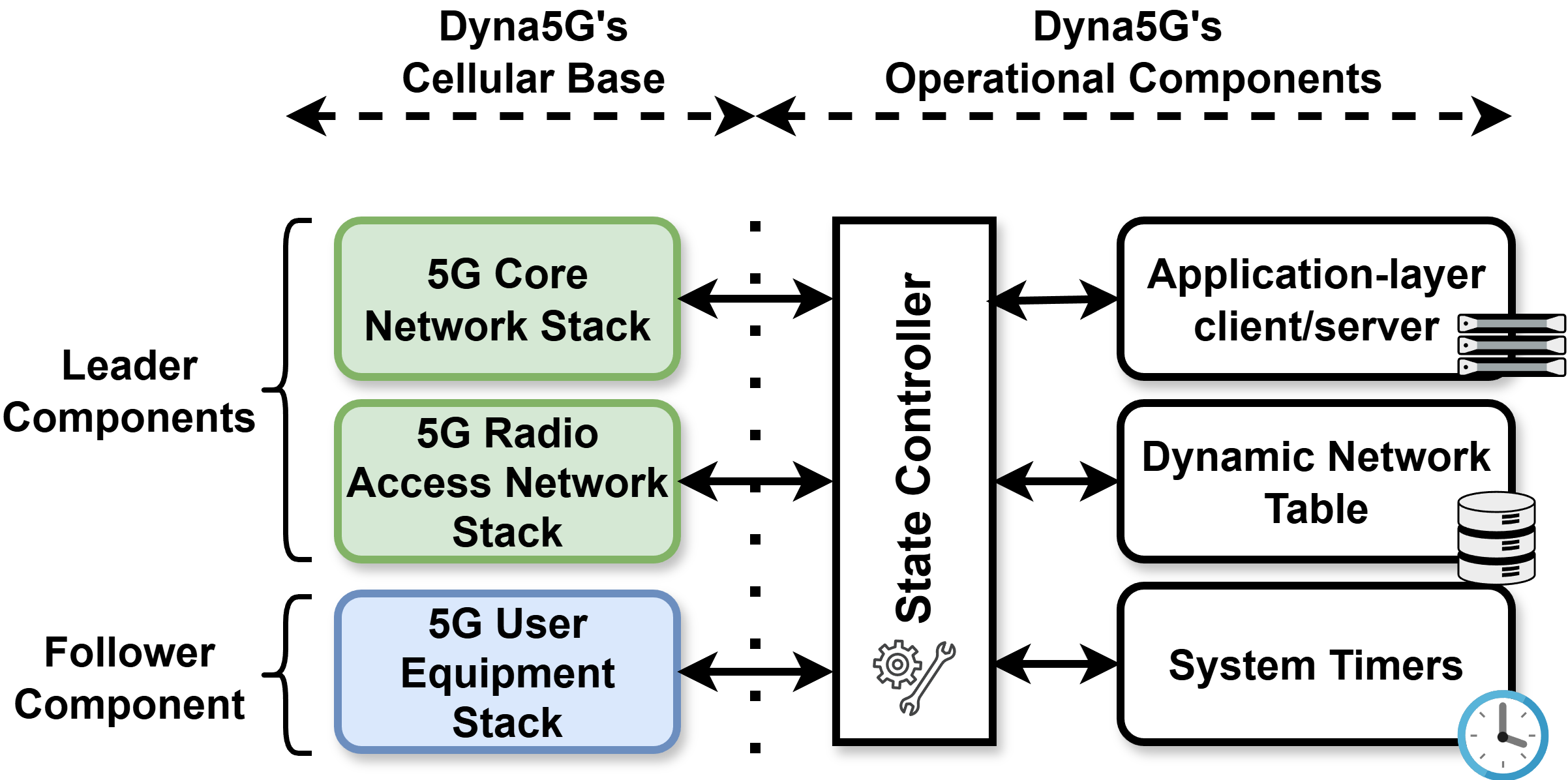}
    \caption{Components placed at each machine. The leader enables the 5G Core and RAN (in green), while the follower the User Equipment (in blue).}
    \Description{}
    \label{fig:components}
\end{figure}

\subsection{Overview}

Figure~\ref{fig:components} shows the main components of each machine, activated according to its current role.

\noindent \textbf{$\bullet$ State Controller.} The state controller contains the Finite State Machine (FSM) logic and it interacts with the other components to make decisions about the network. It enables/disables these components, retrieves important cellular-level, and controls the Non-Access Stratum (NAS) and Radio Resource Control (RRC) transmissions. 

\noindent \textbf{$\bullet$ Dynamic Network Table.} It contains the necessary information (\eg, Session IPs, status) for machine identification, routing and reorganization. The controller updates the table regularly to reflect the latest network state.

\noindent \textbf{$\bullet$ Systems Timers.} They coordinate and drive core control-plane operations, such as performance evaluation and leader selection, and are synchronized across nodes.

\noindent \textbf{$\bullet$ Application Layer Server/Client.} This client and server component is necessary for the exchange of application layer messages between machines. These messages are specific to the dynamic network's management, organization, control, and potentially mission-related instructions and information. 

\noindent \textbf{$\bullet$ UE Stack.} To enable communication between machines, the stack provides standard 5G UE functionality for control-plane and user-plane connectivity, with minor instrumentation added for logging and state reporting to the controller.

\noindent \textbf{$\bullet$ Core \& RAN Stacks.} Activated only on the leader, the stacks provide the infrastructure necessary to initiate and orchestrate the 5G Core and RAN functions to establish the 5G network, with lightweight adaptations for state reporting to the controller.

\subsection{The State Controller}

\begin{figure}[!t]
    \centering
    \includegraphics[width=0.7\columnwidth]{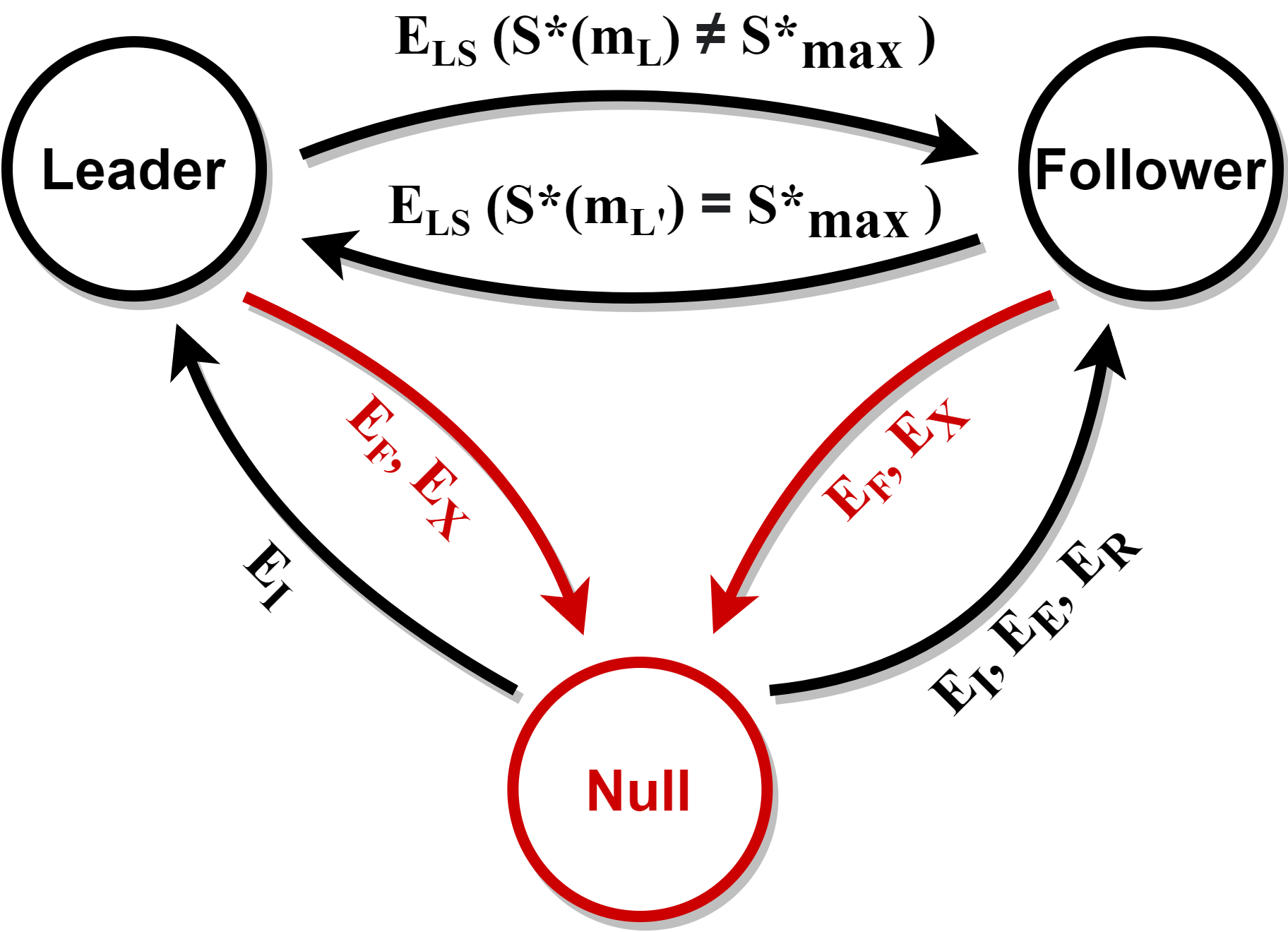}
    \caption{State diagram for all machines in the network. Arrows denote the \textit{Event} triggering each state change.}
    \Description{}
    \label{fig:state-diagram}
\end{figure}

We define the Finite-State Machine (FSM) for the State Controller (Figure~\ref{fig:state-diagram} shows the states and transitions) starting with the set of machines (denoting the nodes) as M = $\{m_1, m_2,$ $\dots, m_N\}$, where \( M \) is the set of all \( N \) machines in the network. Each machine can be in one of the states: \textit{Null}, \textit{Leader}, or \textit{Follower}, \ie, $S(m_i) \in$ $\{\text{Null}, \text{Leader},$ $\text{Follower}\}$, where \( S(m_i) \) represents the state of machine \( m_i \). The ``Follower'' state triggers the machine to function as a UE, while the ``Leader'' state enables the machine to take on the roles of both the Core and RAN. The ``Null'' state indicates that the machine is currently not active in any role. Each machine also has an associated performance score denoted by \( S^{*} \). Specifically, \( S^{*}(m_i) \) is the score of machine \( m_i \).

At the very beginning, all machines start in the \textit{Null} state: \( \forall m_i \in M, S(m_i) = \text{Null} \). At the \texttt{Initialization Phase}, one randomly transitions to the \textit{Leader} state (with individual probability $P(m_i) = 1/N$) and the rest to the \textit{Follower} state. We generally follow the rule that the machine with the highest accepted performance score becomes the leader, \ie, $S(m_L) = \text{Leader}$, \text{where } $m_L = \text{argmax}_{m_i \in M} S^{*}(m_i) = S^{*}_{max}$. Whereas, all other machines become followers denoted as: \( \forall m_i \in M \setminus \{m_L\}, S(m_i) = \text{Follower} \). Machines exchange performance scores throughout the network's lifetime (see Section~\ref{network-organization}). These performance scores are used to identify the next most capable leader. Therefore, for any subsequent leader selection process that potentially entails transition from the current leader to the next (if the next has achieved the best score), we define the next leader as \( m_{L'} \) fulfilling the \( S^{*}(m_{L'}) > S^{*}(m_L) \) and \( S^{*}(m_{L'}) = S^{*}_{max} \). Consequently, the old leader becomes a follower $S(m_L) = \text{Follower}$, while also $S(m_{L'}) = \text{Leader}$.

Additionally, when the current leader fails or exits, the machine with the next best score becomes the leader instantly as $S(m_{L'}) = \text{Leader}$, with  $m_{L'} = \text{argmax}_{m_i \in M \setminus \{m_L\}} S^{*}(m_i)$. As a result, the old Leader \( m_L \) becomes a follower if it reconnects to the network again, $S(m_L) = \text{Follower}$, otherwise it remains in the \textit{Null} state $S(m_L) = \text{Null}$. Followers that exit or fail are handled similarly. Furthermore, when either normally entering the network or as part of the recovery process, the machine always joins as a follower ($S(m_i) = \text{Follower}$) to not disrupt the network status. Table~\ref{tab:fsm_transitions} displays the actions, conditions, events, and FSM transitions.

\subsection{Cellular Stack Integration} \label{cellular-integration}

The integration of 5G cellular stacks is essential for reliable machine communication, forming the foundation for the application layer. In 5G networks~\cite{3gpp.23.501, 3gpp.23.502, 3gpp.38.523, 3gpp.38.401}, the RAN handles wireless communication with UEs via base stations, while the core manages essential network functions. During PDU session establishment, the UE sets up user-plane tunneling, obtains IP addresses, and enables data exchange at the application layer. To support the dynamic M2M network, 5G functions must be adapted for traffic routing, specifically the GPRS Tunneling Protocol for the User Plane with the IP flows to route traffic back to the network and to the target UEs. Open-source solutions like Open5GS~\cite{open5GS24} can support these features with proper configurations. Additionally, the UPF and UEs must share the same IP subnet, with the UPF assigned the X.X.X.1 address as the leader (i.e., Core). Section~\ref{discussion} in the Appendix contains additional information regarding the connectivity and its security establishment.

Because \ourname's orchestration logic is built strictly around 3GPP roles, it operates as a modular layer rather than a custom MAC/PHY overlay. It monitors standardized UE RRC/NAS states (e.g., RRC\_IDLE, RRC\_INACTIVE, RRC\_CO\-NNECTED and registration/session status) and uses them as health signals for role transitions and recovery (see Section~\ref{failure-recovery}). It directly manages gNB and 5GC functions (AMF/SMF/UPF), preserves user-plane reachability via UPF/PDU-session anchoring and tunneling, and retrieves protocol-specific parameters for configurations (e.g., PLMN/TAC, gNB/cell identity, PDU-session/IP allocation) and reorganization (e.g., triggering controlled detach/attach, restarting gNB/AMF/UPF services). \ourname does not modify any 3GPP PHY/MAC/RRC/NAS protocol.


\subsection{Application Layer Setup} \label{application-layer}

This layer is an important part of this network as it incorporates the necessary network messages, such as the \texttt{Performance Report} and \texttt{Heartbeat Notification} (specified in Section~\ref{messages}), which are used in critical network operations (detailed in Section~\ref{phases} of Appendix) and are controlled by the state controller. The communication between machines is possible with each machine hosting one server for reception and one client for transmission. We chose HTTP(S) because it is widely supported, simple to implement, and payload-agnostic, which makes it easy to integrate with existing tools. This layer is crucial because it enables the actual functionalities and services that end-machines interact with, and is transparent to the underlying cellular stack.  However, the application layer is established after the 5G setup, since it relies on the lower network layers for successful communication. 

\section{Phases of Network Operation} \label{phases}
Having defined the FSM and how \ourname hooks into the 3GPP cellular stack, we now detail how the components orchestrating the network: from initial bootstrapping and leader election, through continuous health monitoring and dynamic role switching, to the automated state reconstruction triggered when a node fails.

\subsection{Network Initialization} \label{initialization}

\noindent \textbf{Step 1: Preparation.} When machines are powered on, the network undergoes its initial setup. Besides configuring the application layer and cellular infrastructure, each machine must determine its role before communication can begin. One machine must act as the leader, while the others operate as followers. Since no mutually established leader exists at boot time, each machine initially considers itself a potential leader, and a random selection designates the initial leader.

Upon activation, each machine imports its configuration, including its ID and initial performance score. One machine is assigned a score of 100\% and temporarily assumes the initial leader role, while the remaining machines are initialized as UEs. The leader instantiates the 5G Core and RAN stacks and verifies their interfaces, whereas followers initialize only the UE stack. Each follower then connects to the 5G network by completing RRC connection setup, Authentication and Key Agreement (AKA), and PDU Session establishment, after which it validates the assigned PDU Session IP address for inter-machine communication. The leader obtains its IP address during UPF initialization. All machines are configured within the same IP subnet. Each machine also initializes a \textit{network} table and \textit{heartbeat} records, initially containing only its own information: Machine ID (obtained from the configurations), Device ID (which is the IMEI from the configurations), IP address, coordinates (from GPS), role, and score. Once IP addresses and local tables are available, each machine starts its application-layer server and client. Algorithm~\ref{alg:network-preparation} in the Appendix summarizes these machine steps.

\noindent \textbf{Step 2: Discovery \& Association.} When a follower connects, it sends an \texttt{Entry Notification} to the leader and waits for the \texttt{Entry Notification Reply} before actually joining the network. This step synchronizes the machines by sharing the current network state, including timers, events, and active machine information. The follower extracts the heartbeat, performance-evaluation, and leader-selection timers, along with active machine details (\eg, session IPs and IDs), and updates its internal table. The leader, which always listens for \texttt{Entry Notification} messages, retrieves the relevant network-state information, replies to the joining follower with an \texttt{Entry Notification Reply}, and forwards the update to the other followers for table synchronization. The process concludes with the initialization of threads on each machine for the \textit{Heartbeat}, \textit{Performance Evaluation}, \textit{Leader Selection}, and \textit{Failure Recovery} tasks. At this point, all machines share a synchronized view of the network structure and functions. The leader, however, remains temporary until the next timer-driven evaluation cycle.

\noindent \textbf{Step 3: Post-Association.} After association, all machines listen for broadcast and unicast messages, update their local records, and execute scheduled organizational procedures. Heartbeat, performance-evaluation, leader-selection, and failure-recovery tasks run continuously throughout the network lifetime, alongside mission-related communication.

Machines joining after initialization must complete the association phase to advertise their presence and retrieve the current network state. The same applies to machines that later attempt to re-enter. Such machines always join as followers (see Table~\ref{tab:fsm_transitions}) and advertise a zero score in the \texttt{Entry Notification}, ensuring temporary low privilege and preserving network integrity. This allows them to participate in network procedures without disrupting the mission, while remaining eligible for future leadership. If a machine reconnects after failure, it sends an \texttt{Entry Notification} with a ``reconnection'' cause. Machines may also leave the network voluntarily. To support orderly departure, an \texttt{Exit Notification} informs the other machines to update their tables, remove the departing node from organizational procedures, and terminate related functions (\eg, missions or leader selection). A machine that leaves without sending this message is treated as faulty, in which case failure-detection and recovery mechanisms are applied. In emergencies, a leader may also issue an \texttt{Exit Notification} with cause ``failure alert'' to trigger proactive mitigation.

\noindent \textbf{$\bullet$ Tables \& Timers.} The \textit{Network Table} example presented in Table~\ref{tab:network_table}, is an essential dynamic component of this network's information system. This table integrates diverse but interrelated data points for each machine within the network. For every machine identified by its unique \textit{Machine ID}, the table lists the \textit{Device ID} using the IMEI number, which ensures each drone can be precisely correlated with its cellular module. The \textit{Session IP} address denotes the network connection point for data exchange as instructed by the PDU Session Establishment. The (latest reported) \textit{GPS Coordinates} provide real-time spatial positioning, crucial for navigation, tracking and performance evaluation purposes. The current \textit{Network Role} clearly distinguishes between the `Leader' and the `Followers'. Lastly, the (latest reported) \textit{Performance Score} quantitatively measures each machine's operational effectiveness. In addition to the network table, each machine collects the timestamp of the received heartbeats from others in dynamic \textit{Heartbeat records}. Together, these information constructs a comprehensive picture of the network's functioning, machine status, and collective performance.

\begin{table}[!t]
    \caption{Network Table Example}
    \centering
    \begin{tabular}{|c|c|c|c|c|c|}
        \hline
        \textbf{ID} & \textbf{IMEI} & \textbf{IP} & \textbf{GPS} & \textbf{Role} & 
        \textbf{Score} \\
        \hline
        A1 & 900....0009 & 10.45.0.1 & $X_{1}/Y_{1}/Z_{1}$ & 1 (L) & 85.6 \\
        B2 & 900....0008 & 10.45.0.2 & $X_{2}/Y_{2}/Z_{2}$ & 0 (F) & 72.3 \\
        C3 & 900....0007 & 10.45.0.3 & $X_{3}/Y_{3}/Z_{3}$ & 0 (F) & 67.8 \\
        D4 & 900....0006 & 10.45.0.4 & $X_{4}/Y_{4}/Z_{4}$ & 0 (F) & 63.2 \\
        \hline
    \end{tabular}
    \label{tab:network_table}
\end{table}

The network uses three timers for the heartbeat, performance evaluation, and leader selection, each with intervals based on network requirements. These are shared with new machines via the leader's \texttt{Entry Notification Reply}, along with the time remaining for each process cycle. The leader acquires them from the configurations. Each timer starts at the association phase and remains functional as long as the network is operational. The timers are customizable and depends on the mission and network needs.

\begin{enumerate}

\item \textit{Heartbeat Timer} (\(T_{heartbeat}\)). Runs most frequently to support rapid failure detection and timely dissemination of spatial updates.


\item \textit{Performance Evaluation Timer} (\(T_{performance}\)). Triggers self-diagnostics (e.g., performance assessment) and score sharing for leader selection, and runs less frequently than heartbeat checks to reduce overhead.


\item \textit{Leader Selection Timer} (\(T_{selection}\)). Governs periodic leader selection and reorganization, and runs least frequently to avoid instability, such as ping-pong effects. Under emergencies, it is bypassed for immediate reset.

\end{enumerate}

\subsection{Network Self-Organization} \label{network-organization}

The reorganization aims at maintaining proper network structure, ensuring uninterrupted performance of application layer activities, and determining the most `capable' leader. Figure~\ref{fig:reorganization-processes} illustrates two machines during a reorganization phase, consisting of three simultaneous key phases:

\begin{figure}[!t]
    \centering
    \includegraphics[width=1.\columnwidth]{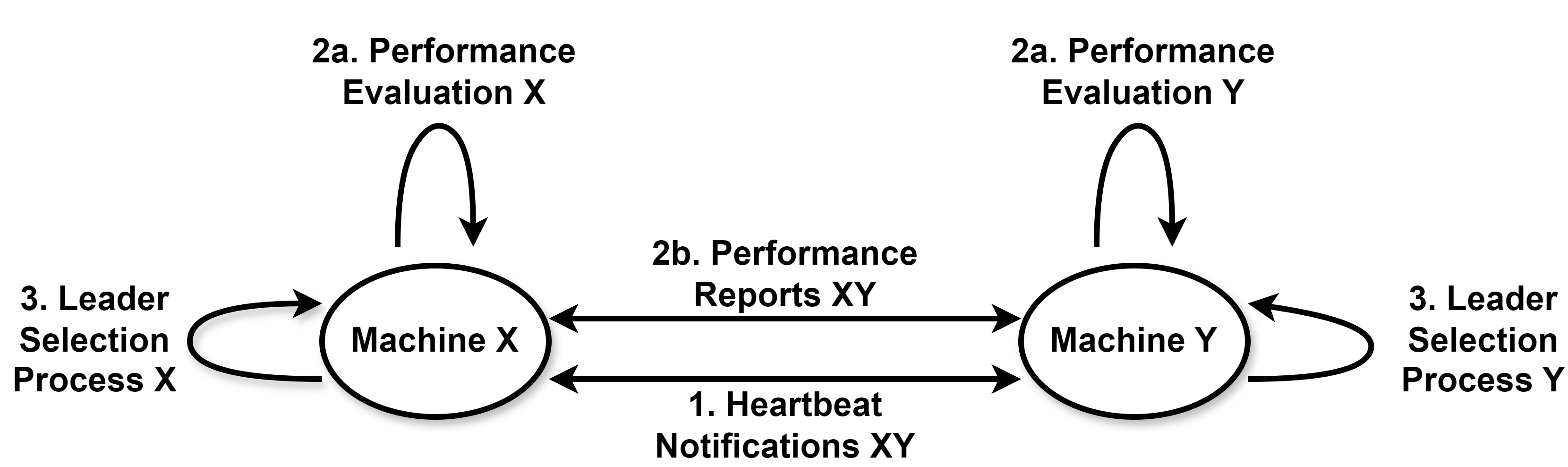}
    \caption{Operations between two machines, exchanging heartbeat notifications (1), conducting performance evaluations (2a), sharing performance reports (2b), and completing the leader selection process (3).}
    \Description{}
    \label{fig:reorganization-processes}
\end{figure}

\noindent \textbf{$\bullet$ Heartbeat Process.} The heartbeat process is executed every time the $T_{heartbeat}$ expires. Each machine is responsible for sending and listening for \texttt{Heartbeat Notification} messages. They contain essential identification, current coordinates, status and timestamp data and are sent to every other active machine. On the receiving end, machines persistently listen for incoming heartbeat messages from their counterparts, updating the ``last seen" timestamp in the heartbeat records, after the message validation. Next, the included coordinates are also updated in the network table.

Heartbeats are also utilized to monitor whether a machine is `alive' or not in the network, but always in accordance with cellular health checks (via cross verification). The timestamp plays a pivotal role in the monitoring phase, wherein each entity oversees the temporal gaps between the current time and the ``last seen" timestamp of every other machine in the network. Failure recovery is triggered if this temporal gap surpasses a specified (customizable) timeout threshold, indicating a potential issue or failure with the corresponding machine. Network entities update their network records to exclude the faulty machine from all operations and communications in case of a faulty follower. The network continues to remain functional without any disruptions. Faulty leaders can either be identified via missing/late heartbeats or when there is no reception of messages (indicating cellular-level faults too, like in routing and PDU sessions). In such a scenario, the handling mechanisms trigger cellular verifications to determine the error and whether a leader selection is eventually required, thereby ensuring the integrity of the network even amidst unforeseen operational issues.

\begin{figure}[t]
    \centering
    \includegraphics[width=0.9\columnwidth]{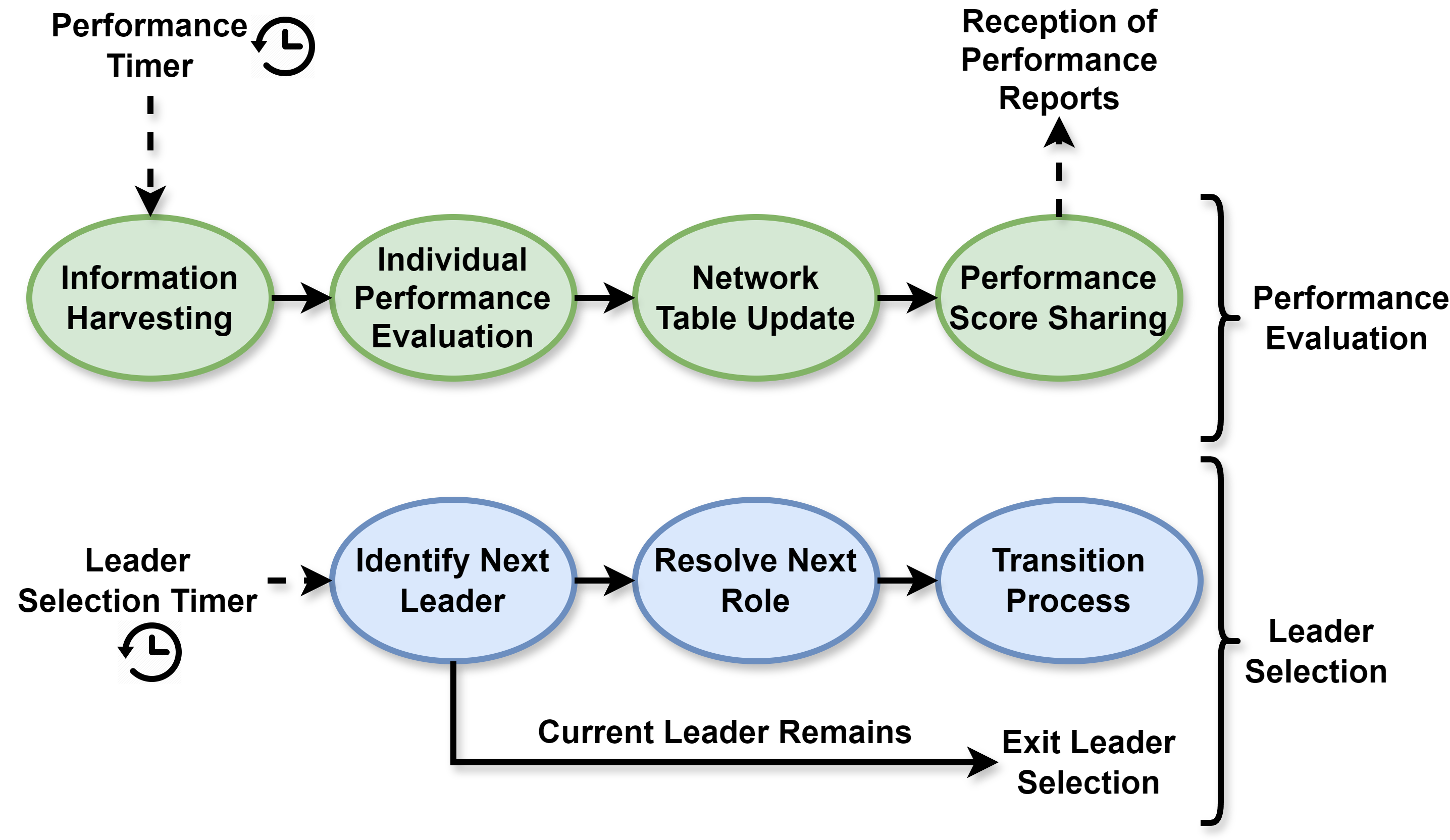}
    \caption{Performance Evaluation \& Leader Selection.}
    \Description{}
    \label{fig:components-processes}
\end{figure}

\noindent\textbf{$\bullet$ Performance Evaluation.} The evaluation begins when $T_{performance}$ expires on all machines. Each machine individually assesses its performance, updates its network table, and shares the results via \texttt{Performance Report} messages to ensure all tables stay updated. Collecting performance reports from other machines in the network eventually leads to updating each machine's network table. Figure~\ref{fig:components-processes} illustrates the components of the Performance Evaluation.

Each machine in the network has unique attributes directly influencing its capabilities and performance. These encompass critical aspects like processing power, memory capacity, battery life, communication range, etc. In \ourname, we consider (1) Proximity to the Network's Center Mass, and (2) Computational/System Capacity, as the primary scoring metrics, which can be adapted based on network requirements and scenarios. \textit{The selected features and algorithms are presented as a proof-of-concept to demonstrate a fundamental scoring method compatible with this network design and implementation.} This design provides the flexibility to integrate any type of algorithm and corresponding threshold adjustments based on the specific mission and its requirements.

\begin{figure}[!t]
    \centering
    \includegraphics[width=0.9\columnwidth]{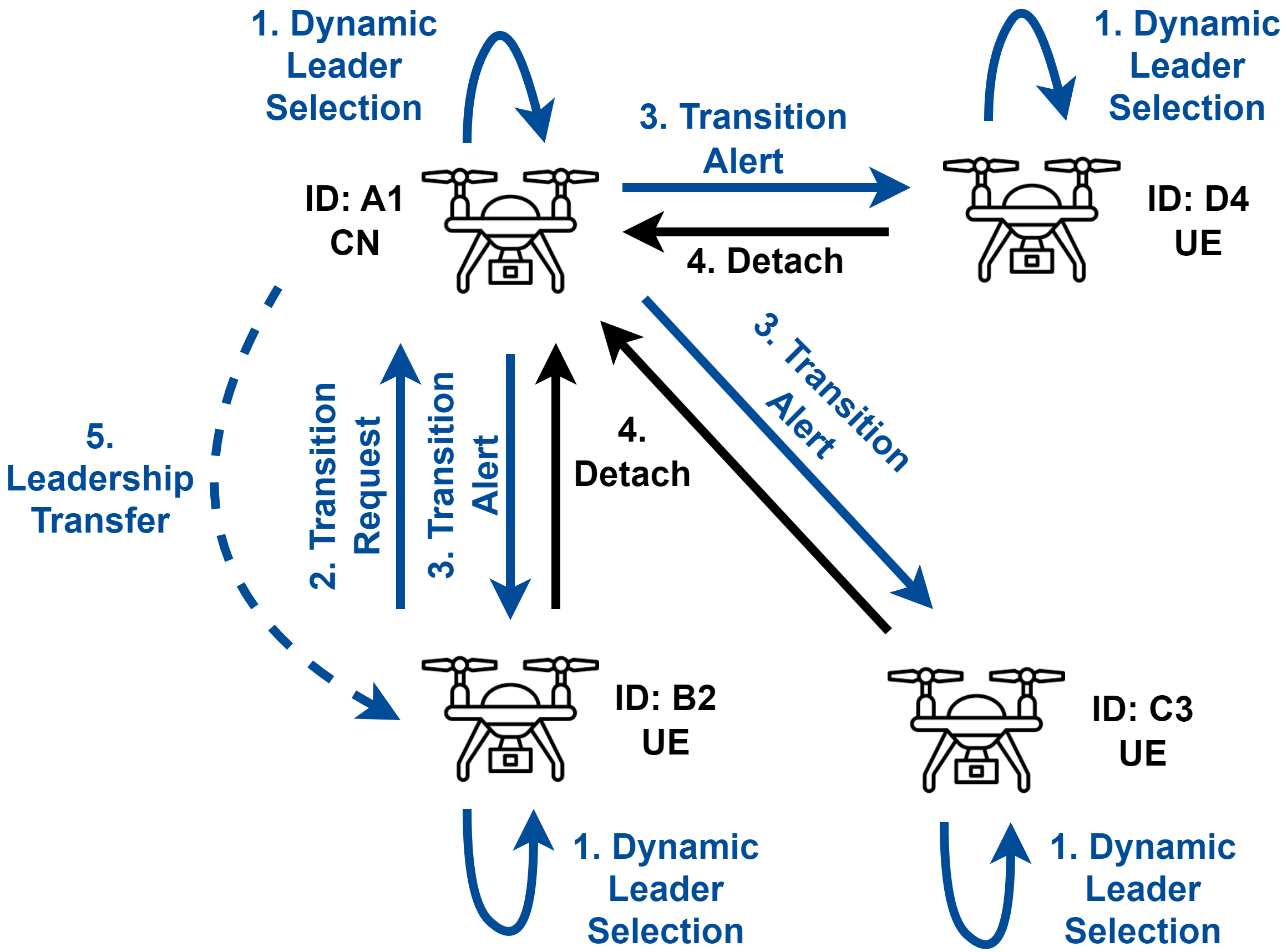}
    \caption{Leader transition process. After all the steps are complete, the network reorganizes around Drone B2 as the new leader. The blue and black arrows are the state-level and cellular-level procedures, respectively.}
    \Description{}
    \label{fig:drone-selection}
\end{figure}

\noindent\textit{Spatial Proximity Score (Sp).} This metric assesses a machine's proximity to the center of mass (COM) of all machines. Lower distances to the COM result in higher scores. The concept is to favor machines closer to all, facilitating effective communication with the leader. Machines at the network's far edges are unsuitable leaders due to inefficient communication. Mathematically, the formula is represented as \(Sp(i) = \frac{100}{1 + a \cdot dist(P(i), COM_{x,y,z})}\), where COM is the center of mass for each dimension (For example, for X, we have \(COM_x \&= \frac{1}{N} \sum_{k=1}^{N} X_k\)), \(dist(\cdot, \cdot)\) is the Euclidean distance between the two input vectors for each dimension, $a$ is the intensity factor ($0 < a < 1$), and \(P(i)\) represents the machine \(i\)'s position in the 3D space.

\noindent\textit{Computational Capacity Score (Cc).} This metric evaluates the machine’s hardware performance based on memory (M), battery (B), and processing power (P). To handle Core and RAN responsibilities, machines must meet minimum resource thresholds (\(\text{iff } M > m\_{\text{thres}} \text{, } B > b\_{\text{thres}}, P > p\_{\text{thres}}\)). Entities with optimal energy reserves can better support the network. The formula \(Cc(i) = (M + B + P)/3\) ensures candidates meet operational standards with \(Cc(i) = 0\) if the conditions are unmet. ``M, B, P" are percentages of available memory (GB), power (W), and processor resources (GHz).

\noindent\textit{Aggregated Performance Score.} The score is computed from two factors as \(S(i) = (Sp(i) + Cc(i))/2\). Each machine then updates its score in the network table, while roles remain unchanged until leader selection is triggered or the leader changes, and disseminates the new value using a \texttt{Performance Report} message. Upon receiving a report, a machine validates it before updating its local table.

\noindent \textbf{$\bullet$ Leader Selection.} At the expiration of $T_{selection}$ or in an emergency, machines identify a potential next leader by retrieving scores from the table, and performing \(i^* = \arg\max_{i \in M} S(i)\), where each machine \(i \in M\) has a corresponding final score \(S(i)\), and (\(i^*\)) is the machine with the maximum score. Once identified, and as long as the maximum score exceeds the predefined offset ($Score_{offset}$), all network tables are updated to reflect the new roles. This offset ensures the top score \(S(i^*)\) is sufficiently higher than the second-best, avoiding ping-pong effects. If the leader remains the same, no change is needed. Figure~\ref{fig:components-processes} displays the integral parts of the selection.

Next, the transition process starts with the candidate machine sending a \texttt{Transition Request} to the current leader. The current leader verifies the request to ensure: (1) its legitimacy, (2) the machine possesses the top score, and (3) the smooth initiation of the leader selection. Once these conditions are met, the current leader broadcasts the \texttt{Transition Alert} to all machines initiating the transition process. Otherwise, it responds with \texttt{Transition Failure}, and the network continues as it is. Once the message \texttt{Transition Alert} is received, all followers disconnect, terminate all network processes, and suspend the application layer. The leader becomes a follower (UE), and the candidate becomes the leader (Core/RAN). The rest remain followers, but reconnect to the new leader. Key elements, such as tables, security context, temporary identifications, and resources (\eg, PDU IP address) are released during the disconnection and subsequently renewed by the incoming leader. It should be noted that this role transition is not a 3GPP RRC handover between cells. RRC handovers and their parameters (measurement events, candidate-cell lists, or A3/TTT/hysteresis) remain unchanged and are not affected by Dyna-5G.

\begin{figure}[!t]
    \centering
    \includegraphics[width=0.9\columnwidth]{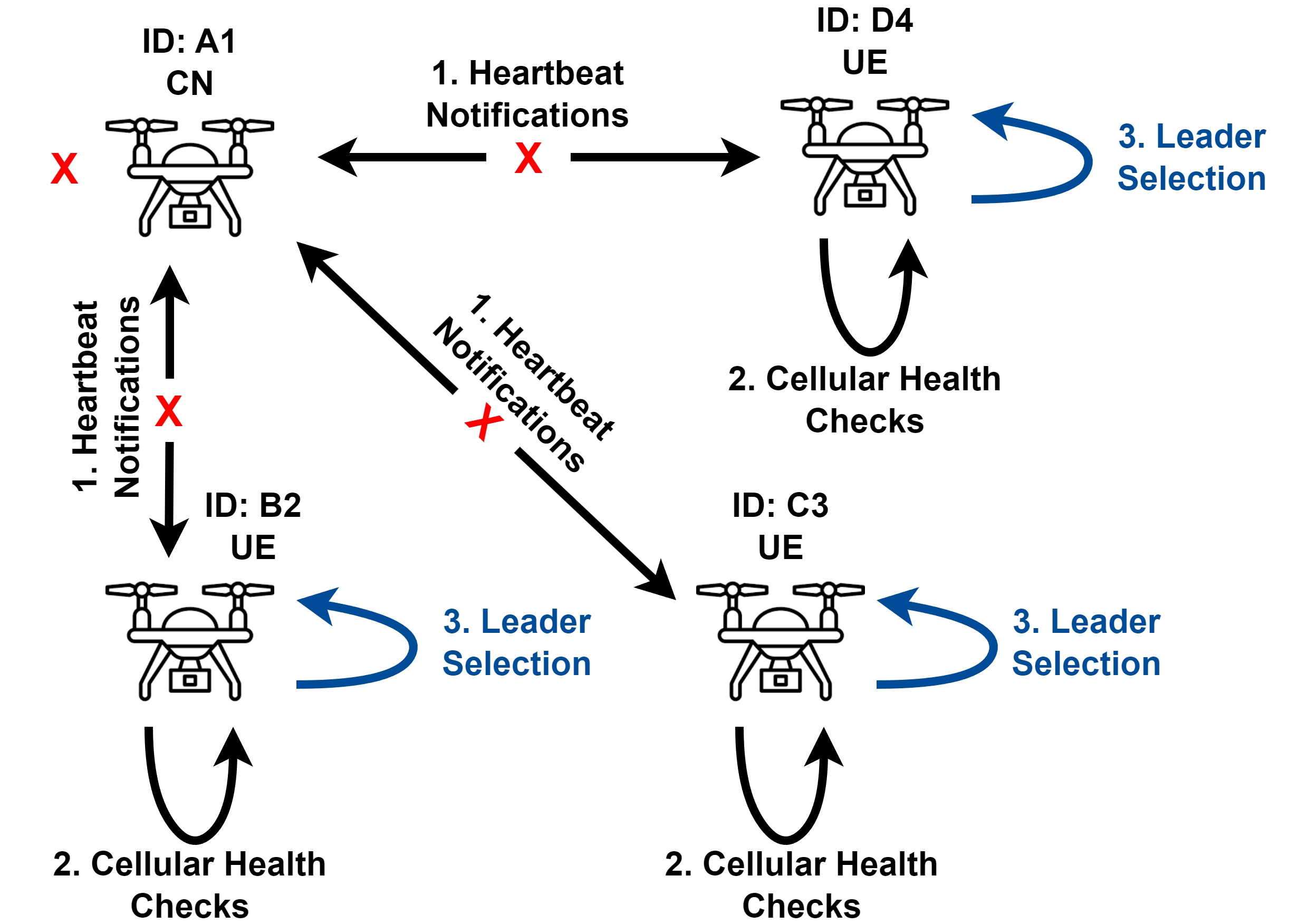}
    \caption{Drone failure recovery. Upon leader collapse (A1), drones detect heartbeat failures and perform cellular-level checks to confirm the leader's status.}
    \Description{}
    \label{fig:drone-recovery}
\end{figure}

\subsection{Failure Recovery Mechanisms} \label{failure-recovery}


These mechanisms detect non-operational machines, with particular emphasis on leader failures. Follower failures result in removal from the network tables and related processes, whereas leader failures trigger immediate emergency leader selection, bypassing timers to accelerate reorganization and state updates. Failure detection relies on heartbeats (see Section~\ref{network-organization}), with missing or delayed signals indicating potential failure, and then on cellular-level checks to confirm that the leader has indeed ceased functioning.
Abrupt crashes, where no \texttt{Exit Notification} is sent, are the true test of this mechanism: the system must detect the failure through missed heartbeats and cellular health checks, then autonomously coordinate the full teardown of the failed leader's state and re-establishment of the network under a new one.



\noindent \textbf{$\bullet$ Cellular-level Health Checks.} These mechanisms detect loss of RAN (gNB) or 5GC functionality by monitoring the progress in the UE RRC/NAS state machines and gNB/5GC service liveness. Each follower continuously observes: (i) PHY synchronization and cell discovery (successful detection/decoding of the 5G Synchronization Signal (SS) and acquisition of MIB/SIB), (ii) RRC connection progress (RRC\_IDLE to RRC\_CONNECTED), and (iii) NAS procedure completion (Registration and PDU Session Establishment). Failures are declared when progress stalls beyond a bounded timeout and/or exceeds a small number of retries. For instance, we flag a core failure when RRC connectivity exists but NAS procedures repeatedly fail to complete (e.g., Registration Request or PDU Session Establishment with no accept/acknowledgment). Additionally, we treat persistent PHY pipeline errors reported by the stack (e.g., RX/TX overflows/underflows) as a degraded-link condition.

When a failure is detected, the node triggers leader reselection using the most recent synchronized network tables (see Section~\ref{network-organization}). The highest-ranked candidate instantiates the leader's gNB+5GC services, advertises the serving cell parameters (e.g., PLMN/TAC/cell ID), and followers perform a controlled detach/attach to reconnect. If the original leader remains alive (e.g., transient desynchronization), followers instead resynchronize to the same cell by reacquiring the SS and continuing normal procedures. Figure~\ref{fig:drone-recovery} illustrates the resulting recovery flow, where leader selection is immediate and recovery primarily requires UE reconnection rather than a full role transition, unlike Figure~\ref{fig:drone-selection}.

\noindent \textbf{$\bullet$ Proactive Measures.} Continuous health monitoring of the leader can check vital metrics like processing power, memory, battery life, cellular status, and responsiveness. Significant deviations or unresponsiveness trigger an \texttt{Exit Notification} from the current leader with a ``failure alert" cause, prompting leader selection and safe exit. Upon receiving this, machines assign a new leader, preventing hard failures and enabling proactive reorganization.

\noindent \textbf{$\bullet$ Handling Failed Message Deliveries.} Devices always verify message reception before initiating processes at timer expiration. Frequent dissemination of key messages (e.g., performance reports) reduces the risk of missed updates. Missing heartbeats, sent every couple of seconds, are also detected and recorded, providing sufficient updates. Heartbeat thresholds can be adjusted based on mission conditions.

\section{Evaluation and Results}
Our evaluation focuses on the feasibility and overhead of dynamic 5G role switching on commodity hardware. Specifically, we quantify the control-plane overhead of the distributed FSM, the baseline system and network capacity available for mission payloads, and the end-to-end latency of automated state reconstruction during various missions (with and without leader failures). We organize our results into three categories: (1) \textit{System Performance}, (2) \textit{Network Performance}, and (3) \textit{Mission Performance}.

\begin{figure}[!t]
    \centering
    \includegraphics[width=\columnwidth]{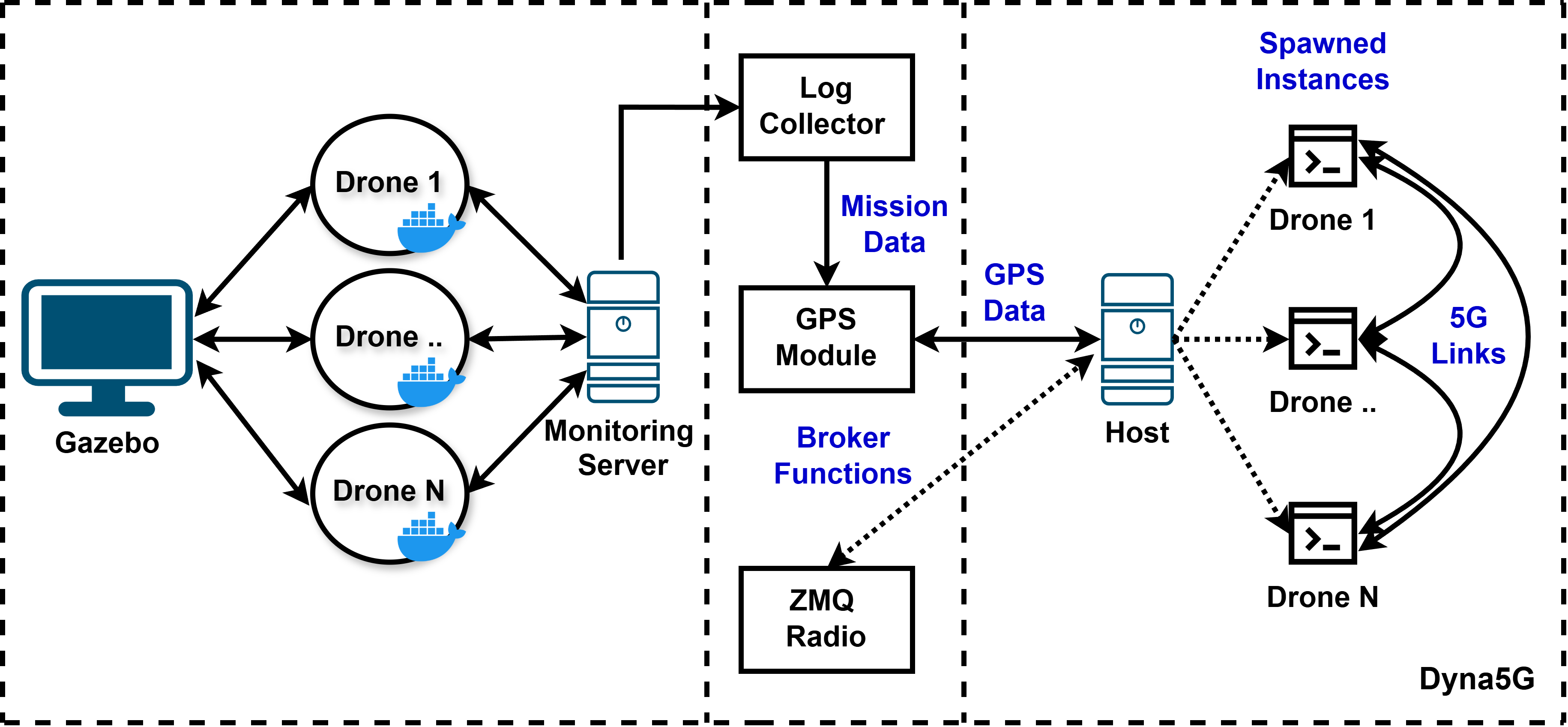}
    \caption{Experimental ecosystem of Dyna5G.}
    \Description{}
    \label{fig:setup}
\end{figure}

We use a custom robotic mission simulator to synthesize missions for swarms of up to 10 drones. We cap the swarm size at 10 because larger swarms make mission generation and trace export computationally intensive and less reliable on our current hardware (although, Dyna5G is \textit{not} limited to 10 instances). However, this is not an inherent limitation of the approach, as swarm evaluations up to 10--20 UAV scales are common in the literature (e.g.,~\cite{kushleyev2013swarm,park2023dlsc,adajania2023amswarm}), with broader surveys also summarizing similar swarm-scales~\cite{coppola2020survey}.

The mission simulator is not part of Dyna5G itself, meaning it is only used to generate mobility and event traces (per-drone positions and mission events) at a 1~second cadence, incorporating Gazebo and ArduPilot. These traces are then imported into Dyna5G at runtime. For the 5G network, we utilize Open5GS~\cite{open5GS24} for the 5G core and srsRAN~\cite{srsran24} for the 5G RAN and UE. All experiments are executed on a commodity laptop (Lenovo Thinkpad, 32~GB RAM, Intel i7-12700H, 1~TB SSD, Ubuntu~20.04). Nonetheless, we emphasize that Dyna5G introduces no additional hardware beyond the RF modules already required by the baseline 5G setup, as all changes are software-based (e.g., orchestration/coordination and telemetry). Figure~\ref{fig:setup} illustrates a high-level overview of the experimental ecosystem: a custom robotic mission simulator generating GPS logs of drone coordinates during the mission, our intermediate functions and the Dyna5G environment. The log collector aggregates all necessary information from the executed missions, which is then processed. The logs input to the GPS module, a custom Python program for synchronization with Dyna5G. The host also initiates and maintains the proper ZeroMQ instance for drone connection over cellular, and then spawns drone instances with the correct configurations.~\footnote{As our baseline, we utilized srsRAN's guideline for ZeroMQ~\cite{srsRAN24:zmq}.} Section~\ref{discussion} in the Appendix provides more information regarding the setup.

\subsection{System Performance}
The main system performance metrics are the CPU, memory and disk utilization. We collected measurements every second for 30~seconds during the missions. The measurements were aggregated and processed into a CVS file. The processing involves the average value calculations for each swarm size. We also include the baseline behavior of the system without Dyna5G running in order to determine the real resource consumption of the architecture. Table~\ref{tab:system} shows the system's metrics and values for several swarm sizes (One drone instance is denoted as `Base'). For each size, we then have four categories in percentages; CPU user, CPU System, Memory used, and Disk utilization.

On our resource-constrained setup, Dyna5G’s functions remain lightweight. More capable servers can reduce these averages even further. Although we have not tuned for efficiency, straightforward optimizations, such as thread affinity, batching of control messages, and allocator pooling for message buffers, can likely reduce CPU and memory overhead even further. It should be noted that since Dyna5G’s processes run on the server/ground host in our setup, the reported CPU, memory, and disk overhead is incurred on that host. For a UAV, the single-instance footprint is feasible for several classes, like RB5 platforms~\cite{qualcomm23:5g-platform}.

\noindent\textbf{$\bullet$ CPU.} Average CPU utilization (user+system) scales with swarm size and remains within headroom on a single host. Baseline average is $\approx$3.68\% (2.60\% user, 1.08\% system). At 10 drones, the mean rises to $\approx$47.2\% (29.4\% user, 17.8\% system), a net $\approx$43.5\% above baseline. Intermediate sizes fall between $\approx$10.8 \% (2 drones) and $\approx$17.6\% (8 drones). This increase is consistent with the number of active threads (leader selection, performance evaluation, heartbeat, failure recovery, and the per-node server) per instance and the additional protocol work (GPS playback + GNU Radio/ZeroMQ). Importantly, even the largest configuration leaves substantial CPU headroom for telemetry and application workloads.

\noindent\textbf{$\bullet$ Memory.} Memory use grows with the number of instances but stays modest for the evaluated scales. Baseline averages $\approx$3.93 GB, and the 10-drone run averages $\approx$14.16 GB (a +10.23 GB delta). Extrapolating at roughly $\approx$1 GB per additional active instance (upper bound from our runs), a 20-drone experiment would stay on the order of $\approx$24 GB total, fitting comfortably within common 32–64 GB lab servers. We observed no paging or allocator churn, resident set sizes remained stable, and the memory consumption did not affect the mission execution.

\noindent\textbf{$\bullet$ Disk.} The reported disk utilization remains essentially flat around 3.1–3.5\% across all configurations, indicating no storage pressure from logging or temporary files.

\noindent\textbf{$\bullet$ Power.} We note that the observed incremental power draw did not exceed $\approx$25 Watts over baseline during the heaviest experiments on our testbed (which is within the payload power available on many enterprise UAV platforms, e.g., Matrice-class~\cite{dji.m300rtk.manual}), consistent with the CPU and memory figures above. In operational deployments, follower-side energy can be further reduced using standard NR power-saving mechanisms such as connected-mode DRX (configured via RRC/MAC) and wake-up receiver/signal concepts~\cite{3gpp.38.321,3gpp.38.331,3gpp.38.774}. For the leader (gNB) role, 3GPP explicitly studies RAN/network energy saving features (e.g., BS sleep and related mechanisms) that can reduce base-station energy consumption~\cite{3gpp.38.864}. Overall, we expect the energy overhead to be manageable, since it can be optimized, and the leader election is energy-aware (see Section~\ref{network-organization}) rotating the leader to avoid sustained drain on any single drone.

\begin{table}[!t]
  \centering
  \caption{System metrics for various swarm sizes.}
  \label{tab:system}
  \setlength{\tabcolsep}{3pt} 
  \renewcommand{\arraystretch}{1.1}  
  \rowcolors{3}{gray!6}{white}            
  \begin{tabularx}{\columnwidth}{
      >{\raggedright\arraybackslash}p{0.25\columnwidth} 
      >{\centering\arraybackslash}X                    
      >{\centering\arraybackslash}X                   
      >{\centering\arraybackslash}X                 
      >{\centering\arraybackslash}X                   
      >{\centering\arraybackslash}X                   
      >{\centering\arraybackslash}X                  
  }
    \toprule
    \rowcolor{gray!14}
    \makecell[l]{\textbf{Metrics}} &
    \textbf{Base} & \textbf{2} & \textbf{4} & \textbf{5} & \textbf{8} & \textbf{10} \\
    \midrule
    CPU User (\%) & 2.6 & 6.9 & 10.9 & 8.62 & 11.8 & 29.4 \\
    CPU Sys. (\%) & 1.08 & 3.89 & 5.75 & 5.22 & 5.83 & 17.8 \\
    Memory (MB) & 3,933 & 7,478 & 9,173 & 10,013 & 9,838 & 14,158 \\
    Disk (\%) & 3.28 & 3.52 & 3.09 & 3.37 & 3.3 & 3.3 \\
    \bottomrule
  \end{tabularx}
  \vspace{0.25em}
  {\footnotesize The values include consumption from the OS, GPS module and GNU-ZMQ.}
\end{table}

\subsection{Network Performance}

Our goal is to show that Dyna5G can execute missions reliably while keeping network overhead modest across swarm sizes. We ran hundreds of trials for 2, 4, 5, 8, 10 drones and collected: (1) Full-interface packet captures for control traffic, (2) Bidirectional \texttt{iPerf}~\cite{iperf3} tests (TCP/UDP), and (3) ICMP/TCP RTT probes. Traces were analyzed with \texttt{tshark}~\cite{wireshark.tshark.man} to produce time-resolved Bitrate and packet-rate series, as well as per-flow reliability indicators (e.g., retransmissions).

\begin{table}[!t]
  \centering
  \caption{Network performance metrics for various swarm sizes for control-plane traffic.}
  \label{tab:network}
  \setlength{\tabcolsep}{4pt}           
  \renewcommand{\arraystretch}{1.1}    
  \rowcolors{3}{gray!6}{white}         
  \begin{tabularx}{\columnwidth}{
      >{\raggedright\arraybackslash}p{0.36\columnwidth}  
      >{\centering\arraybackslash}X                    
      >{\centering\arraybackslash}X                    
      >{\centering\arraybackslash}X                   
      >{\centering\arraybackslash}X                   
      >{\centering\arraybackslash}X 
      >{\centering\arraybackslash}X  
  }
    \toprule
    \rowcolor{gray!14}
    \makecell[l]{\textbf{Metrics}} &
    \textbf{2} & \textbf{4} & \textbf{5} & \textbf{8} & \textbf{10} & \textbf{10*} \\
    \midrule
    Data Size (KB) & 69 & 608 & 1,117 & 3,200 & 4,792 & 6,919 \\
    Data Byte Rate (KB/s) & 0.5 & 3.6 & 6.1 & 16 & 34 & 58 \\
    Data Bit Rate (Kbps) & 3.7 & 28 & 49 & 129 & 273 & 470 \\
    Avg Packet Size (B) & 98,93 & 98,41 & 98 & 97,43 & 97,78  & 98 \\
    Avg Packet Rate (p/s) & 4 & 36 & 62 & 166 & 349 & 599 \\
    \bottomrule
  \end{tabularx}
  {\footnotesize 10* corresponds to 10 drones with the fastest transmission possible.}
\end{table}

Across all swarm sizes, control-plane messages remain small (mean packet length $\approx$98 B) and periodic (e.g., heartbeats, performance updates, transition signaling), according to Table~\ref{tab:network}. As the swarm grows, traffic scales as expected for a peer-to-peer control scheme, but the absolute bandwidth remains low. Even in the fastest 10-drone configuration, the average control-plane load is $\approx$0.47 Mb/s, which corresponds to $\approx$0.47\% of a 100 Mb/s 5G bearer and $\approx$4.7\% of a 10 Mb/s LTE bearer. Thus, control traffic is effectively non-intrusive and leaves the vast majority of capacity for telemetry and application data. The gradual increase in packet rate with N is consistent with per-pair periodic exchanges, while the stable mean packet size reflects compact message formats.

\begin{table}[!t]
\caption{iPerf and RTT measurements for 10 drones (fastest transmissions) for application-layer data.}
\centering
\setlength{\tabcolsep}{1pt}
\rowcolors{3}{gray!6}{white}
\begin{tabular}{@{}p{0.42\columnwidth}p{0.53\columnwidth}@{}}
\toprule
\rowcolor{gray!14}
\textbf{Metric} & \textbf{Value} \\ 
\midrule
ICMP latency & $\approx$18.45 ms RTT (20 attempts) \\
ICMP latency (followers) & $\approx$48.26 ms RTT (20 attempts) \\
TCP latency & $\approx$17.77 ms RTT (20 attempts) \\
TCP latency (followers) & $\approx$36.70 ms RTT (20 attempts) \\
\midrule
TCP throughput & $\approx$43.8 MB, 18.3 Mb/s \\
Streams tested & up to 16 \\
Retransmissions & 0\% \\
\midrule
UDP loss @ 16 Mb/s & 0\% \\
UDP jitter @ 16 Mb/s & $\approx$0.779 ms \\
UDP @ 50 Mb/s & saturated link; loss and failures \\
\bottomrule
\end{tabular}
\label{tab:network-2}
\end{table}

Table~\ref{tab:network-2} evaluates \emph{drone payload exchange capabilities}, independently of control signaling. With 10 drones under ``fastest transmissions”, we observe latency in the tens of ms (ICMP/TCP RTT $\approx$18–48~ms depending on leader/follower paths), indicating responsiveness for near-real-time telemetry and coordination. Next, the TCP throughput is in the tens of Mb/s with 0\% retransmissions, supporting reliable bulk transfers. UDP performance sustains $\approx$16 Mb/s loads with 0\% loss and sub-millisecond jitter ($\approx$0.78 ms). Pushing UDP to 50 Mb/s saturates the radio path (expected for a stress test), so practical operating points should target the pre-saturation region (e.g., $\le$ 15–20 Mb/s per active stream or aggregate, depending on concurrency and scheduling).

Our measurements are deliberately conservative with the 5G link using standard QoS and cell/physical layer configurations, default RLC settings, and with ZeroMQ introducing aggregation points and delays absent in a production 5G stack. Despite these \textit{worst-case} choices, the network capabilities allow for dense telemetry data exchange, time-sensitive coordination messages, and routine bulk transfers such as mission logs. Within the pre-saturation region, the drones can also carry moderate compressed video streams (e.g., few Mb/s) alongside telemetry without contention. These results indicate scalability is not a limiting factor, as further optimization would only increase headroom. Additionally, because Dyna-5G acts as an orchestration layer and our RF pipeline is deterministic, PHY/MAC KPIs remain unchanged relative to the srsRAN baseline (i.e., without Dyna-5G), with high CQI (typically 12–15 in our setup), correspondingly high MCS, near-zero DL/UL BLER ($\approx$0–1\%), and negligible HARQ/RLC retransmissions, during normal Dyna5G operations. Overall, the architecture is network-efficient and can support higher-rate application streams.

\subsection{Mission Performance}
We evaluate mission behavior using our custom robotic simulator to synthesize three representative missions across multiple swarm sizes: (M1) \emph{Normal Forward Formation}, where the swarm advances along a common route while maintaining a compact convoy around the leader, (M2) \emph{Area Coverage Formation}, where drones spread out to uniformly cover a target region while preserving connectivity, and (M3) \emph{Leader Drifts and Asymmetric Course}, where the leader is forced onto a deviating trajectory that progressively pulls the formation off-center to test mission adaptations. Each mission runs for $\approx$120~seconds with drone speeds of $\sim$ 2 m/s and above, which is sufficient to trigger steady-state operation, role changes, and potential disturbances. For every mission–swarm configuration we perform repeated executions (hundreds of runs) to identify the general trends and ensure observed effects are not artifacts of a single trace. In addition to nominal runs, we explicitly inject failures to both leader and follower to examine Dyna5G's reorganization under adverse conditions. 

Note that Dyna5G itself imposes no architectural limit on swarm size; the practical ceiling in our testbed arises from the ZeroMQ-based integration and the underlying \texttt{srsRAN} stack's UE instantiation path. For instance, we spawned 12–14 software instances, but reliably attaching all UEs at once requires care in the random access (RACH) procedure to avoid uplink contention (we introduced conservative delays in the connection logic to mitigate this). ZeroMQ introduces aggregation points and latency that are not representative of a production 5G user-plane (also noted by \texttt{srsRAN}), so we retained these conservative settings to ensure repeatability rather than peak performance. Consequently, results in this section too should be viewed as \textit{lower-bound (worst-case)} mission behavior for the control- and data-plane coordination that Dyna5G performs. It should also be stated that ZeroMQ does not affect Dyna5G's orchestration logic.

\noindent \textbf{$\bullet$ Normal Scenarios.} In nominal runs, Dyna5G executes all three missions with stable role management and predictable adaptation. Two parameters dominate steady-state behavior: the fixed hysteresis used to prevent ping pong effects (i.e., Offset from Section~\ref{network-organization}) and the vital intervals (i.e., heartbeat, performance evaluation, leader selection). For all runs, we retained our rule $T_{heartbeat}$ $<$ $T_{performance}$ $<$ $T_{selection}$, which indicates that each process must be completed at least once before the next one executes to retain proper network functionality. We also follow the scoring method of Section~\ref{initialization}.

Regarding the \emph{Offset}, we identified the baseline of 5.5–6.0\% producing correct operation but higher sensitivity, yielding up to 2.5\% leader transitions per mission on average across our traces (all sizes). Increasing the offset to 9.0\% reduced unnecessary leader handovers while preserving responsiveness, lowering the mean for large swarms from $\approx$3.6\% to $\approx$2.7\% and for smaller swarms from $\approx$1.7\% to $\approx$1.2\% transitions per mission. This confirms the expected trade-off: smaller offsets respond more readily to small score fluctuations (risking ping-pong and extra stack work), whereas larger offsets are more conservative and save resources. Additionally, the scoring algorithm should be taken into consideration, which is anyway an adaptive element in our architecture.

Moreover, for the \emph{Intervals}, we discovered that the ``2–4–6'' schedule (heartbeat–performance–leader in seconds) was our empirical baseline for these specific missions. This is because it allowed enough time for drones to perform the associated functions. Values that went beyond that schedule (e.g., ``3–5–7'') resulted in slow network updates, cohesion issues, and missed reelections and failure detections. Table~\ref{tab:swarm-missions} in the Appendix presents experimental measurements for the baseline with offset 5.5\%. Specifically, it reports the number of transitions, median of score differences for transitions, and leader sequence, for all combinations of missions-sizes. 

In contrast, to tighten reaction time, we also adopted the ``1–2–4'' schedule (fastest transmissions), which increased the input freshness, the updates-evaluations frequency and network reliability without exhibiting instability in our runs. Shortening the leader-selection interval specifically raises sensitivity, as it is calibrated jointly with the offset to avoid ping-pongs, without affecting crucial transitions. Consequently, with the ``1–2–4'' cadence and an $\approx$9\% offset, missions completed without missed adaptations while avoiding superfluous role changes, with leader management remaining fast and stable, limiting transition frequency and preserving resources (unlike some unnecessary transitions for the configurations in Table~\ref{tab:swarm-missions}, especially for large swarms). Regardless, these vital parameters for mission and network efficiency are user-configurable and tunable based on the mission and requirements, thus making Dyna5G adaptive.

Finally, we identified that Core, RAN and UE initialization are negligible for the mission timing overhead (i.e., a few milliseconds in our setup). The dominant delays observed in nominal transitions stem from ZeroMQ's synchronization and traffic handling rather than radio stack start-up and Dyna5G's functions. Although the control logic itself supports rapid reconfiguration, we measured that ZeroMQ can add extra seconds to the pipeline (e.g., wait, aggregate input from all UEs and then send them to gNodeB) that are not typical under normal RF conditions. These results are grounded in our repeated experiments across missions and swarm sizes. Besides, the system was eventually able to execute the missions successfully with various configuration setups and no practical constraints, while retaining its scalability.

\noindent\textbf{$\bullet$ Failure Scenarios.} We evaluate the reorganization robustness by injecting single-node crashes into the three missions at controlled times (e.g., at 60 or 80~seconds) for swarms of 5, 8, and 10. Crashes are emulated at the GPS layer by returning the sentinel coordinate $(-99999,-99999,-99999)$ for the failed drone. Concurrently, the dedicated cellular stack shuts down and the node is treated as inactive, as it misses consecutive heartbeats beyond the configured threshold and fails the cellular health checks. 

Failure handling proceeds deterministically. First, the cellular checks (e.g., RRC and NAS) indicated a failure, and the subsequent missed heartbeat(s) triggered the failed node removal from the heartbeat map and network table. Next, the emergency leader-selection routine ran immediately (using the same scoring logic as in nominal case) and elected the highest-scoring remaining node. The transition mechanism reset the network. Normal metrics/parameters, i.e., \emph{Heartbeat}, \emph{Performance}, \emph{Leader-Selection} intervals and the \emph{Ping-Pong Offset} remain unchanged and unaffected. Across all tested sizes and missions, the emergency leader selection was triggered as intended, the failed node was excised from state, and connectivity for the remaining drones was restored without additional operator intervention. Figure~\ref{fig:mission-failure} in the Appendix shows a failed leader in the coverage scenario for 5 drones. The leader crashed and did not reach the expected position, while the rest continued uninterrupted.

\begin{figure}[!t]
    \centering
    \includegraphics[width=0.85\columnwidth]{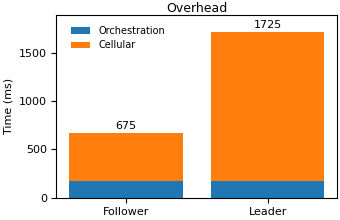}
    \caption{Network transition overhead in our setup.}
    \Description{}
    \label{fig:overhead}
\end{figure}

We studied the process executions and the code during the experiments, and measured the time durations. In all cases, the (emergency) leader-election decision completed within tens of milliseconds. Excluding deliberate stabilization waits/sleeps (added to support ZeroMQ queueing/batching and transport delays), Dyna5G's orchestration overhead for reorganization \textit{under steady-state conditions in our setup} (no issues with threading, I/O, OS) includes (as per Figure~\ref{fig:overhead}): (1) only around 175~ms timing overhead purely from our orchestration logic, and (2) an approximate cellular (re)initialization overhead of 500~ms (follower) and 1,550~ms (leader). Together, these components yield an overall failure recovery path of approximately 2.5~s in our setup, with around 86\% of the time due to stack-dependent cellular procedures. Thus, under both normal and faulty operations, Dyna5G’s logic remains lightweight and does not constitute a bottleneck in the recovery.

\section{Conclusion}

We built and evaluated \ourname to show that a fleet of commodity machines can autonomously operate a 5G SA network, recover from leader failure without human intervention, and complete missions successfully. Across hundreds of trials with up to 10 drones, control-plane overhead averaged $\approx$0.47~Mb/s, leader crashes were recovered in $\approx$2.5~s, with \ourname's own orchestration logic contributing only $\approx$175~ms of that. Beyond the numbers, the key finding was that the hardest part of 5G failure recovery, coordinating distributed state teardown and re-establishment across Core, RAN, and all UEs simultaneously, can be fully automated without modifying any 3GPP protocol. We hope \ourname lowers the barrier for adopting cellular 5G in autonomous M2M fleets where resilience remains a main obstacle today.


\bibliographystyle{ACM-Reference-Format}
\bibliography{main}


\appendix

\section{Discussion} \label{discussion} 

\noindent\textbf{ZeroMQ Challenges.} ZeroMQ imbues some limitations in the experimentation, even though it allows us to test our system on a large scale. Apart from the delays and additional resource exhaustion, another obstacle is that the aggregation points for the uplink expect inputs from all instances before they proceed. Nonetheless, we are consistent with srsRAN guidelines, and therefore treat ZeroMQ as a conservative, non-optimized integration layer, very useful for repeatable experiments, but not representative of Dyna5G’s best-case performance (can be improved further). In the future, we are planning to investigate the ZeroMQ pipeline further.

\noindent\textbf{Scalability Considerations.} For this study we prioritized a software-only path that guaranteed reproducibility, fine-grained control and integration with simulation environments. Hardware-in-the-loop with USRPs is valuable for RF realism but impractical for the large scales we target. A 10-drone swarm would require \emph{at least} 10 hosts and 10 radios (plus cabling, power, synchronization, and spectrum isolation), quickly exceeding lab resources and complicating repeatability. Even with sufficient hardware, multi-node orchestration (clocking, calibration, and over-the-air coexistence) becomes the bottleneck rather than the Dyna5G mechanisms under test. Our goal here is to stress the control and data-plane at scale, create repeatable experiments and \emph{evaluate Dyna5G's system logic} thoroughly.

Furthermore, emulation with Colosseum~\cite{bonati21:colosseum} can approximate large RF environments, but in practice it proved unsuitable for our present campaign. The platform was operationally complex, we encountered intermittent connection/setup errors, and critically, \texttt{srsRAN~5G} is not currently supported in our required configuration. Scaling to large swarms also demands reserving many instances concurrently, increasing the chance of resource contention and experiment breakage. These issues jeopardize the throughput of repeated trials, the tight control we need over failure injection and timing, and above all, the precision in system, network and mission measurements.

\begin{figure}[!t]
     \centering
     \includegraphics[width=0.8\columnwidth]{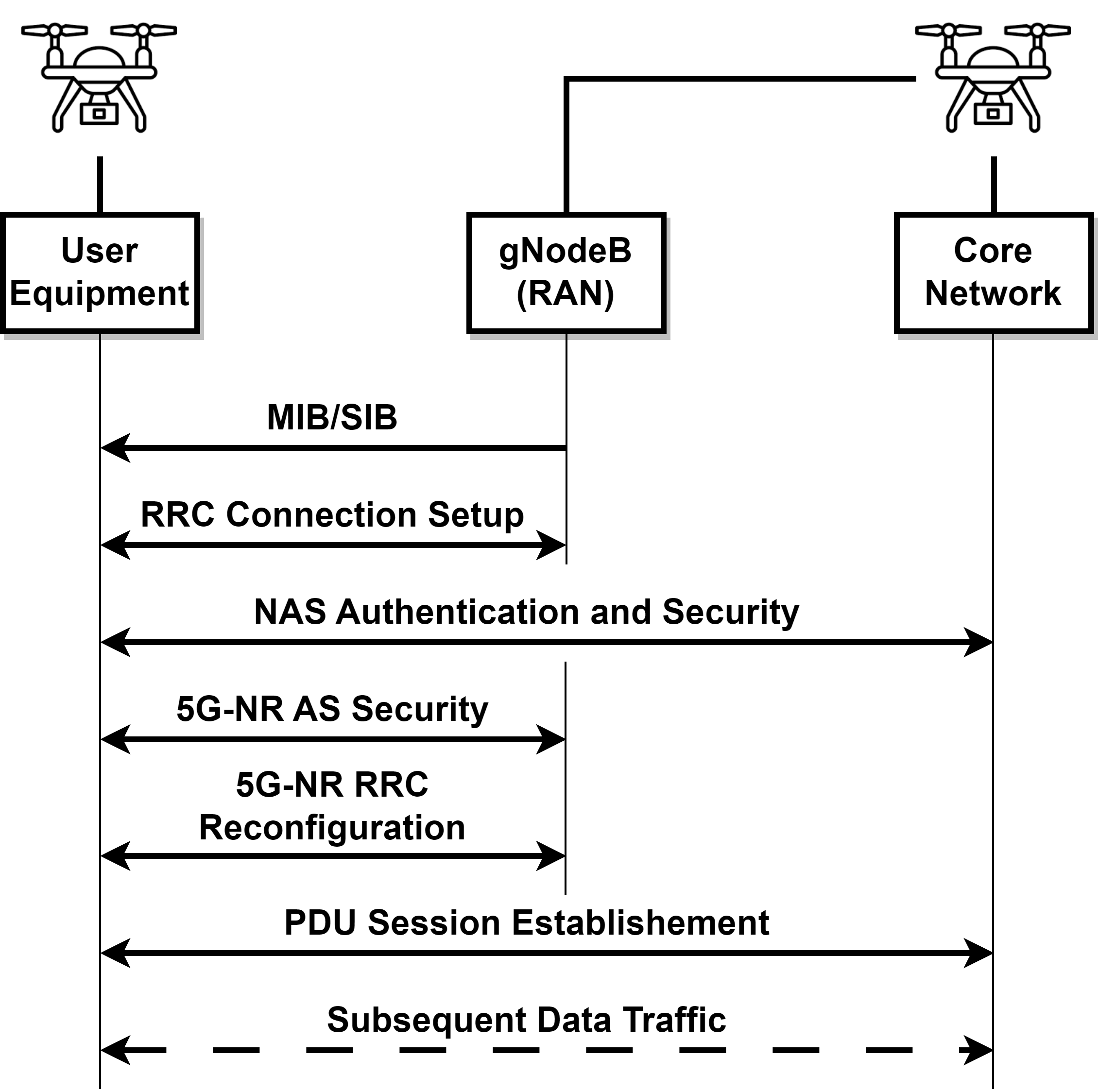}
     \caption{UE-Drone establishing connection with the network. The RAN and Core are hosted by another drone, which has been selected as network leader.}
     \label{fig:drone-security}
     \Description{}
 \end{figure}

\noindent\textbf{Security Considerations.} Our focus in this work is on dynamic topology management for a trusted fleet, not on designing new security/cryptographic protocol or framework. We assume that each drone is provisioned with a valid USIM and attaches to a private 5G core using standard 5G AKA, integrity protection, and encryption. Only UEs that successfully authenticate under this private PLMN obtain a PDU session and can participate in Dyna-5G, meaning all application-layer control messages (Entry Notifications, Heartbeats, Performance Reports) are exchanged over these protected tunnels. Sensitive configuration and identity material (e.g., SIM credentials) are stored in the local database that is protected at rest (e.g., encrypted and access-controlled), and are never logged or exposed in clear text. When a leader fails and a new leader takes over, the affected UEs re-attach and establish fresh security context with the core, releasing temporary identifiers and keys associated with the old leader, as described in our failure-recovery mechanisms. This ensures that the 5G security guarantees are maintained throughout the lifetime of the network and mission. Figure~\ref{fig:drone-security} shows the 5G security establishment and communication that drones perform in Dyna5G, by utilizing the UE, RAN and Core stacks. 

Beyond assuming a trusted fleet, we believe that compromises due to potential leader/follower failures can be further mitigated (in addition to the aforementioned encryption at rest) by: (i) protecting sensitive subscriber material with hardware-backed secure storage (e.g., TPM/TEE-backed keys bound to measured boot~\cite{tcg.tpm2.part1}), (ii) using split-knowledge methods (e.g., $k$-of-$n$ threshold sharing~\cite{10.1145/359168.359176}) so that no single drone can reconstruct long-term credentials, and (iii) restricting leader eligibility to a small set of hardened ``leader-capable'' nodes (e.g., drones equipped with secure elements/TPM and remote attestation), while allowing the rest of the swarm to operate only with ephemeral session context.

\noindent\textbf{Trust and Bootstrap.} With the established security measures and baselines, newly joining or rejoining drones always enter as followers (as already stated). After a successful attach, the drone issues an \texttt{Entry Notification}, is inserted into the network table with its reported metrics, and remains in the follower role until it has completed at least one full performance-evaluation interval. This ensures that a newly connected or potentially unstable node cannot immediately self-elect as leader or disrupt an ongoing mission, meaning it must first demonstrate stable performance and be selected through the standard scoring and selection cycle. This indicates that network has the capabilities to enforce robust access control, prioritization and authorization. In this design we assume trust between nodes, also due to the established cellular security standards, but additional mechanisms (such as certificates, consistency checks, or reputation systems) can be implemented as well. These can handle extreme cases of compromised drones (e.g., falsifying scores, flooding heartbeats, or selectively dropping traffic). Although these are orthogonal to the design we evaluate here, we leave this as future work, as they are out of scope for this study.

\noindent\textbf{Implementation of Leader-Selection Policy.} Dyna5G is implemented primarily in C++ for performance, and the leader-selection algorithm is isolated in a single C++ module. This module exposes a stable interface that maps the drone and mission state to a score. To adopt a different policy, a user only needs to modify the algorithm and recompile; the FSM, timers, message formats, and cellular stack remain unchanged. Because the rest of the system only consumes these scores and the resulting leader identifier, \emph{any} scoring algorithm that produces a total ordering over candidates (from simple heuristics to learned models) can be embedded in this module \emph{without affecting Dyna5G’s core logic}. As already specified, the policy we implement in this paper is a proof of concept. A systematic exploration of alternative algorithms is considered as out of scope.

\section{Conventional Drone Communication Over Cellular} \label{conventional-net}

\begin{figure}[!t]
     \centering
     \includegraphics[width=0.9\columnwidth]{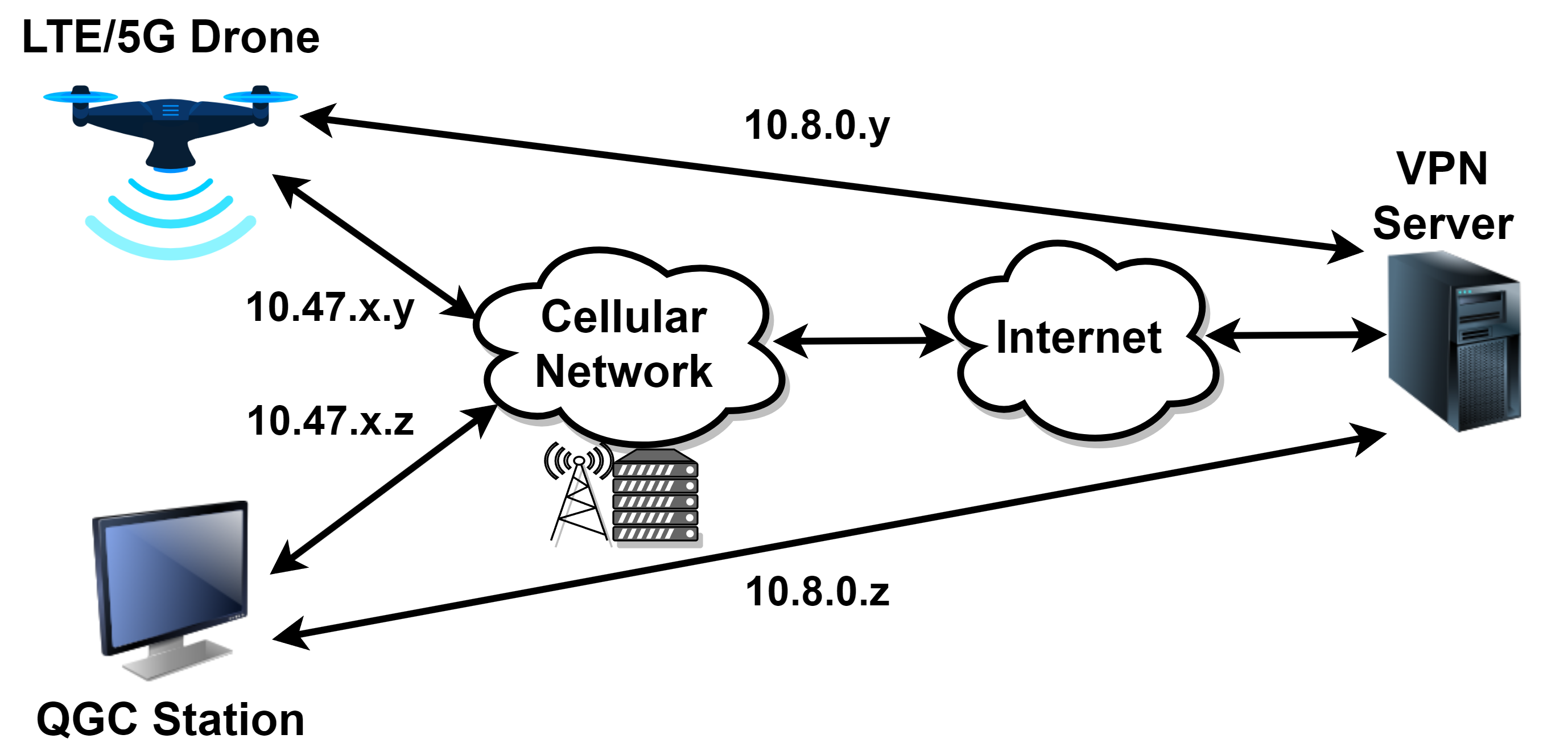}
     \caption{Connection of a LTE/5G drone to the QGC via a commercial-private cellular network~\cite{modalai23:5G-Modem}.}
     \label{fig:drone-connection}
     \Description{}
 \end{figure}

In the realm of LTE/5G-enabled drones, communication between a QGroundControl (QGC) station and a drone predominantly relies on cellular networks, a practical methodology supported by Qualcomm and ModalAI~\cite{modalai23:5G-Modem, qualcomm23:5g-platform}. As depicted in Figure~\ref{fig:drone-connection}, this setup enables drones to be remotely managed via the QGC, facilitated by either private or commercial cellular providers. Initially, both drone and QGC establish a connection to the 5G core network, where they are assigned distinct session IP addresses (\eg, drone: 10.47.x.y, GCS: 10.47.x.z) for internet access.

Subsequently, to ensure secure communication, both entities connect to a Virtual Private Network (VPN) server via their cellular links, receiving additional IP addresses (drone: 10.8.0.y, GCS: 10.8.0.z) from the VPN, which runs server software in the cloud and client software on both drone and GCS. This arrangement facilitates direct communication between the drone and GCS, leveraging unique VPN IP addresses assigned to each. For consistent and secure connections, it is crucial that the drone and GCS maintain static IP addresses upon each VPN connection. This is achieved by allocating unique security certificates to each network endpoint, ensuring that upon connection with its certificate, each device consistently receives the same IP address. Eventually, the communication involves an IP packet encapsulated, with the application-layer data, by another IP packet.

 \begin{figure}[!htb]
  \centering
  \begin{subfigure}{0.23\textwidth}
    \centering
    \includegraphics[width=\linewidth]{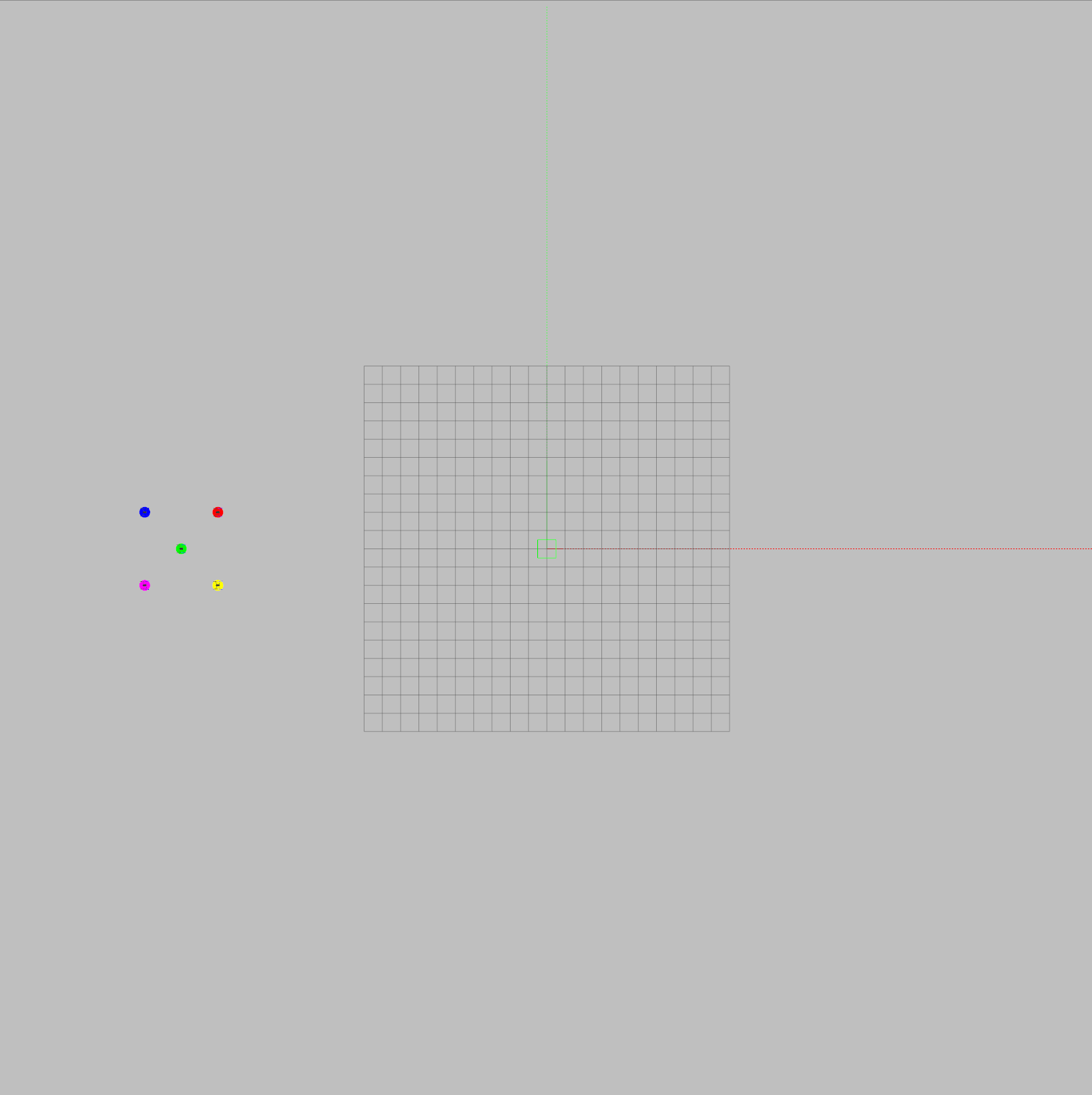}
    \caption{Start of the mission}
    \label{fig:left}
  \end{subfigure}\hfill
  \begin{subfigure}{0.23\textwidth}
    \centering
    \includegraphics[width=\linewidth]{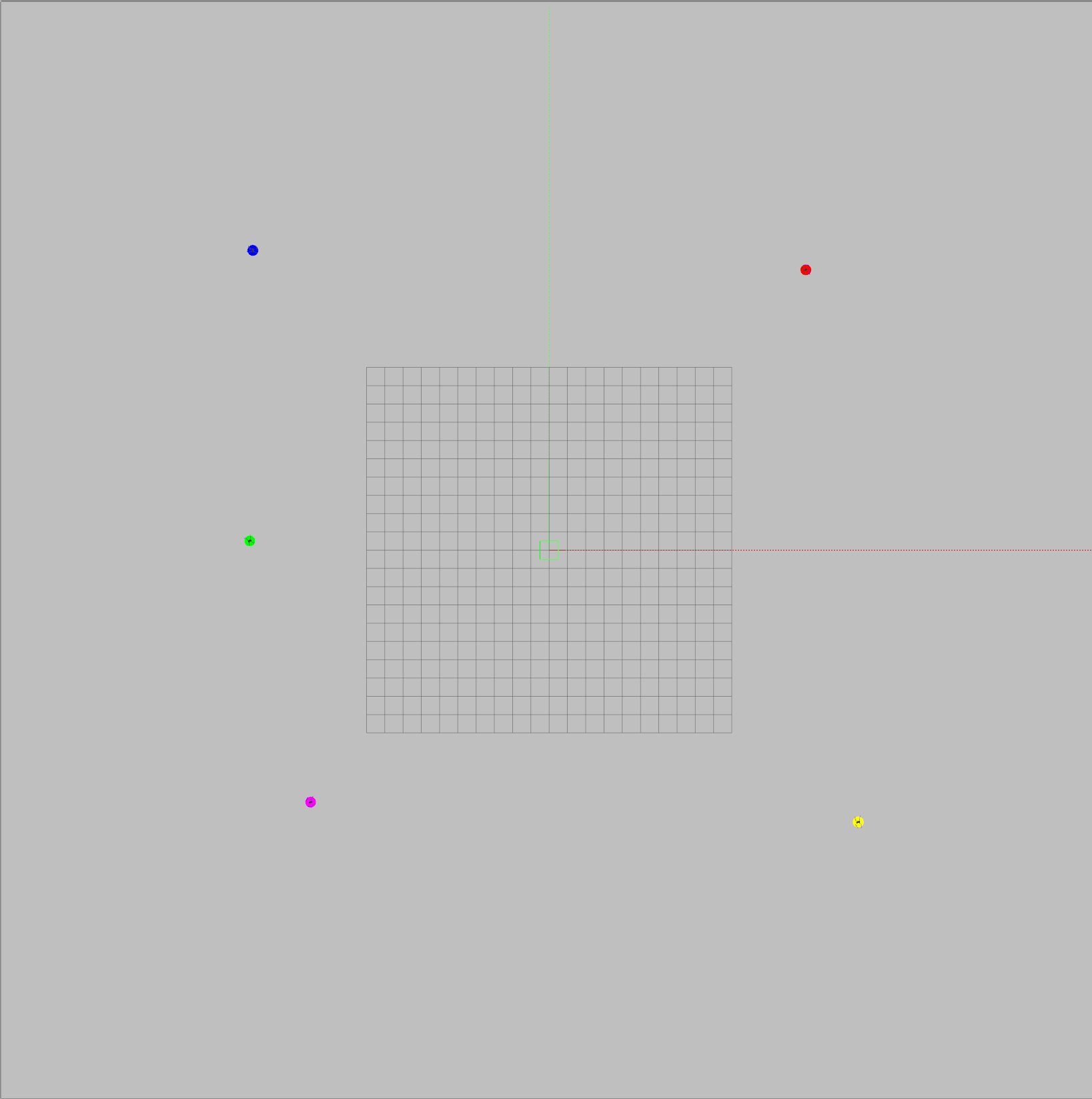}
    \caption{End of the mission}
    \label{fig:right}
  \end{subfigure}
  \caption{Failed leader (in green) during the coverage mission.}
  \label{fig:mission-failure}
  \Description{}
\end{figure}

 \newcommand{\breakseq}[1]{%
  \StrSubstitute{#1}{-}{\allowbreak-}[\temp]\temp
}

\newcommand{\cellmc}[3]{%
  \begin{minipage}[t]{\linewidth}\centering
    \scriptsize \textbf{\#T}: #1\\[0pt]
    \scriptsize \textbf{$\Delta\bar S$}: #2\\[0pt]
    \scriptsize \textbf{Seq}: \breakseq{#3}%
  \end{minipage}%
}

\begin{table}[!ht]
  \centering
  \caption{Metrics by swarm size and mission type for our baseline and offset 5.5\%. All missions were completed successfully.}
  \label{tab:swarm-missions}
  \setlength{\tabcolsep}{1.5pt}     
  \renewcommand{\arraystretch}{1.25} 
  \setlength{\aboverulesep}{0.3ex}    
  \setlength{\belowrulesep}{0.3ex}
  \rowcolors{2}{gray!6}{white}

  \begin{tabularx}{\columnwidth}{
      >{\raggedright\arraybackslash}p{0.15\columnwidth}
      *{6}{>{\centering\arraybackslash}X}
  }
    \toprule
    \rowcolor{gray!14}
    {\textbf{M/S}} & \textbf{S=2} & \textbf{S=4} & \textbf{S=5} & \textbf{S=8} & \textbf{S=10} \\
    \midrule
    M1 & \cellmc{1}{7}{1-2} & \cellmc{3}{8}{1-2-1-4} & \cellmc{2}{15}{1-5-1} & \cellmc{4}{10}{1-2-3-2-1} & \cellmc{3}{8}{1-2-6-1}  \\
    M2 & \cellmc{1}{6}{1-2} & \cellmc{2}{6.5}{1-2-1} & \cellmc{0}{--}{1} & \cellmc{2}{9.5}{1-2-4} & \cellmc{2}{11.5}{1-2-6} \\
    M3 & \cellmc{1}{11}{1-2} & \cellmc{2}{11}{1-2-3} & \cellmc{3}{6}{1-2-3-5} & \cellmc{5}{14}{1-3-2-6-5-2} & \cellmc{6}{9}{1-2-4-6-3-8-4} \\
    \bottomrule
  \end{tabularx}
  {\footnotesize \#T is the number of transitions, $\Delta\bar S$ is the median of the score differences in transitions, $Seq$ is the leader succession, $S$ is the size.}
\end{table}

\begin{algorithm}[!ht]
\caption{Network Initialization (Concise)}
\label{alg:network-preparation}
\SetKwFunction{InitCore}{initiate\_5G\_core\_stack}
\SetKwFunction{InitRAN}{initiate\_RAN\_stack}
\SetKwFunction{GetUPF}{get\_upf\_address}
\SetKwFunction{InitUE}{initiate\_UE\_stack}
\SetKwFunction{GetIP}{get\_IP\_address\_of\_PDU}
\SetKwFunction{Verify}{verify\_connection\_with\_5G\_Core}
\SetKwFunction{CreateTable}{create\_network\_table}
\SetKwFunction{AddEntry}{add\_entry\_to\_network\_table}
\SetKwFunction{SetupServer}{setup\_server}
\SetKwFunction{SetupClient}{setup\_client}

\If{performance\_score == 100}{
    \InitCore{}\;
    sleep()\;
    \InitRAN{}\;
    ip\_address $\gets$ \GetUPF{}\;
    role $\gets$ 1\;
}
\ElseIf{performance\_score $<$ 100}{
    sleep()\;
    \InitUE{}\;
    sleep()\;
    \Verify{}\;
    ip\_address $\gets$ \GetIP{}\;
    role $\gets$ 0\;
}
\CreateTable{}\;
\AddEntry{machine\_id, device\_id, ip\_address, coordinates, role, performance\_score}\;
\SetupServer{ip\_address, 8080}\;
\SetupClient{}\;

\end{algorithm}

\section{Baseline Control-Plane Message Types} \label{messages}

The following are the HTTP messages that have been designed and implemented to support the Dyna5G functionalities and procedures.

 The \textbf{Entry Notification} structure encapsulates:
\begin{itemize}[leftmargin=*,noitemsep,topsep=0pt]
    \item \textit{source machine id} and \textit{destination machine id}: The IDs of the sender and receiver of the message.
    \item \textit{timestamp}: The time at which the message was created.
    \item \textit{imei}: The unique identifier for the device.
    \item \textit{role}: The role of the sender within the network.
    \item \textit{performance}: The current performance metrics of the sender.
    \item \textit{ip address}: The network address of the source machine.
    \item \textit{gps\_x}, \textit{gps\_y}, and \textit{gps\_z}: The geographical coordinates (latitude, longitude, and altitude) of the sender at the time of the message.
    \item \textit{cause}: The reason or cause for the entry notification.
\end{itemize}

 \noindent The \textbf{Entry Notification Reply} structure encapsulates:
\begin{itemize}[leftmargin=*,noitemsep,topsep=0pt]
    \item \textit{source machine id} and \textit{destination machine id}: The IDs of the sender and receiver of the message.
    \item \textit{timestamp}: The creation time of the message.
    \item \textit{selection timer interval}, \textit{heartbeat timer interval}, and \textit{evaluation timer interval}: Timers for leader selection, heartbeat signals, and performance evaluation.
    \item \textit{estimated performance}: The estimated time for the next performance evaluation cycle.
    \item \textit{estimated leader selection}: The estimated time for the next leader selection cycle.
    \item \textit{network table entries}: All entries of the network table representing the current network topology.
\end{itemize}

 \noindent The \textbf{Exit Notification} structure encapsulates:
\begin{itemize}[leftmargin=*,noitemsep,topsep=0pt]
    \item \textit{source machine id} and \textit{destination machine id}: The IDs of the sender and receiver of the message.
    \item \textit{timestamp}: The time at which the message was created.
    \item \textit{role}: The role of the sender within the network.
    \item \textit{cause}: The reason for the exit notification.
\end{itemize}

 \noindent The \textbf{Performance Report} structure encapsulates:
\begin{itemize}[leftmargin=*,noitemsep,topsep=0pt]
    \item \textit{source machine id} and \textit{destination machine id}: The IDs of the sender and receiver of the message.
    \item \textit{timestamp}: The time at which the message was created.
    \item \textit{performance score}: The performance score of the machine.
\end{itemize}

 \noindent The \textbf{Heartbeat Notification} structure encapsulates:
\begin{itemize}[leftmargin=*,noitemsep,topsep=0pt]
    \item \textit{source machine id} and \textit{destination machine id}: The IDs of the sender and receiver of the message.
    \item \textit{timestamp}: The time at which the message was created.
    \item \textit{cellular status}, \textit{vehicle type}, \textit{autopilot}, \textit{base mode}, and \textit{system status}: Operational parameters including cellular status, type of vehicle, autopilot settings, base mode, and system status.
    \item \textit{Vn}, \textit{Ve}, and \textit{Vd}: Velocities in the north, east, and down directions, respectively.
    \item \textit{X}, \textit{Y}, and \textit{Z}: Location coordinates.
    \item \textit{heading}: The direction in which the vehicle is pointed.
\end{itemize}

 \noindent The \textbf{Transition Request} structure encapsulates:
\begin{itemize}[leftmargin=*,noitemsep,topsep=0pt]
    \item \textit{source machine id} and \textit{destination machine id}: The IDs of the sender and receiver of the message.
    \item \textit{timestamp}: The time at which the message was created.
    \item \textit{candidate score}: The score indicating the machine's candidacy for transition.
    \item \textit{cause}: The reason for issuing the transition notification.
    \item \textit{network status} and \textit{transition plan}: The current network status and the proposed steps for the transition.
\end{itemize}

 \noindent The \textbf{Transition Alert} structure encapsulates:
\begin{itemize}[leftmargin=*,noitemsep,topsep=0pt]
    \item \textit{source machine id} and \textit{destination machine id}: The IDs of the sender and receiver of the message.
    \item \textit{timestamp}: The time when the message was created.
    \item \textit{approved candidate id}: The identifier of the approved candidate.
    \item \textit{transition start time}: The scheduled start time for the transition.
    \item \textit{network configuration change}: The adjustments to the network configuration.
\end{itemize}

 \noindent The \textbf{Transition Failure} structure encapsulates:
\begin{itemize}[leftmargin=*,noitemsep,topsep=0pt]
    \item \textit{source machine id} and \textit{destination machine id}: The IDs of the sender and receiver of the message.
    \item \textit{timestamp}: The time when the message was created.
    \item \textit{failure cause}: The reason for the failure.
    \item \textit{retry policy}, \textit{suggestive action}, and \textit{supporting data}: Guidance on subsequent steps, including retry policies, recommended actions, and any additional supporting data.
\end{itemize}

\end{document}